\documentclass[french]{hermes-journal}

\usepackage[utf8]{inputenc}
\usepackage{graphicx}
\usepackage{url}
\newcommand{\slug}{\hbox{\kern1.5pt\vrule width2.5pt height6pt depth1pt\kern1.5pt}}
\def\xskip{\hskip 7pt plus 3pt minus 4pt}
\def\sskip{\vskip 3pt plus 1pt minus 1pt}
\newdimen\algindent
\newif\ifitempar \itempartrue 
\def\algindentset#1{\setbox0\hbox{{\bf #1.\kern.25em}}\algindent=\wd0\relax}
\def\algbegin #1 #2{\algindentset{#21}\alg #1 #2} 
\def\aalgbegin #1 #2{\algindentset{#211}\alg #1 #2} 
\def\alg#1(#2). {\medbreak 
  \noindent{\bf#1}({\it#2\/}).\xskip\ignorespaces}
\def\algstep#1.{\ifitempar\sskip\noindent\else\itempartrue
  \hskip-\parindent\fi
  \hbox to\algindent{\bf\hfil #1.\kern.25em}%
  \hangindent=\algindent\hangafter=1\ignorespaces}
\dn{2015}{18}{2-3}

\firstpagenumber{123}

\title[Ordonnancement d'entités à l'âge des deux web]{Ordonnancement d'entités pour\\la rencontre du web des documents\\et du web des données}

\author[1]{Mazen}{Alsarem}
\author[1]{Pierre-Edouard}{Portier}
\author[1]{Sylvie}{Calabretto}
\author[2]{Harald}{Kosch}

\address{Universit\'{e} de Lyon, CNRS\\
INSA de Lyon, LIRIS, UMR5205, F-69621, France}
{prenom.nom@insa-lyon.fr}
\address{Universit\"{a}t Passau\\
Innstr. 43, 94032 Passau, Germany}
{Prenom.Nom@uni-passau.de}

\resume{Les avancées de l'initiative Linked Open Data (LOD) ont permis de mieux structurer le web des données. Quelques jeux de données servent de centralisateurs (e.g., DBpedia) qui maintiennent les différentes sources de données du LOD liées entre elles.  Ces jeux de données ont également permis le développement de services de détection des entités du \textit{web des données} dans une page du \textit{web des documents} (e.g., DBpedia Spotlight).  Pour permettre l'émergence de nouveaux usages qui combineront les deux webs, nous proposons un algorithme qui ordonne les entités détectées dans une page web en fonction de l'expression d'un besoin d'information. Nous utilisons cet algorithme pour construire un système de génération de snippets sémantiques dont nous montrons expérimentalement l'utilité et l'utilisabilité.}

\abstract{The advances of the Linked Open Data (LOD) initiative are giving rise to a more structured web of data. A few datasets act as hubs (e.g., DBpedia) connecting many other datasets. They also make possible new web services for entity detection inside plain text (e.g., DBpedia Spotlight), thus allowing for new applications that will benefit from a combination of the web of documents and the web of data. To ease the emergence of these new use-cases, we propose a query-biased algorithm for the ranking of entities detected within a web page. Our algorithm combines link analysis with dimensionality reduction. We use crowdsourcing for building a publicly available and reusable dataset on which we compare our algorithm to the state of the art. Finally, we use this algorithm for the construction of semantic snippets for which we evaluate the usability and the usefulness with a crowdsourcing-based approach.}

\motscles{
web des données,
ordonnancement d'entités,
snippets sémantiques.
}


\begin{document}

\maketitle


\section{Introduction}
Dans ce travail, nous présentons l'algorithme LDRANK qui permet d'ordonner les entités d'un graphe issu du LOD (Linked Open Data) en fonction d'un besoin d'information exprimé sous la forme d'une requête formée d'un ensemble de mots clés. L'algorithme LDRANK s'applique à des graphes pour lesquels des données textuelles descriptives peuvent être associées aux n{\oe}uds (c.à.d. aux entités). Nous appliquons cet algorithme à des graphes issus de la détection automatique au sein d'une page web d'entités du LOD. Cette détection d'entités peut être réalisée grâce à des services tels que DBpedia Spotlight~\cite{mendes2011dbpedia} (configurable par le biais de listes blanches et noires qui permettent de filtrer sur les types d'entités tels que définis par la hiérarchie des classes de l'ontologie DBpedia, incluant une phase de désambiguïsation contextuelle, et associant un score de confiance à chaque résultat), AlchemyAPI\footnote{{\scriptsize\url{www.alchemyapi.com}~; \url{www.opencalais.com}~; \url{www.ontos.com}}} (similaire à DBpedia Spotlight, mais faisant référence à différentes sources de données du LOD et comportant donc une étape de résolution des coréférences), OpenCalais\footnotemark[\value{footnote}], SemanticAPI de Ontos\footnotemark[\value{footnote}], etc. Les données textuelles associées aux n{\oe}uds du graphe proviennent alors du résumé DBpedia de l'entité, ainsi que de passages de la page web centrés sur les occurrences de l'entité. Afin de motiver la nécessité de LDRANK, il est nécessaire de rappeler brièvement le rapport entre les approches classiques pour l'ordonnancement de pages web et les approches existantes pour l'ordonnancement d'entités du web des données.

Dans le contexte du \textit{web des documents}, un hyperlien indique une relation entre des informations portées par deux pages web. Bien que de telles relations soient généralement d'une granularité assez grossière, elles forment cependant une composante essentielle des algorithmes d'ordonnancement les plus reconnus (PageRank~\cite{page1999pagerank}, HITS~\cite{kleinberg1999authoritative}, SALSA~\cite{lempel2001salsa}).

Dans le contexte du \textit{web des données}, les liens (matérialisés par des triplets sujet/prédicat/objet) représentent des relations nommées dont les entités sources et cibles sont généralement d'une granularité plus fine que pour le web des documents. La grande majorité des stratégies existantes pour l'ordonnancement d'entités du web des données (voir~\cite{roasurvey} et~\cite{jindal2014review} pour des états de l'art récents) sont fondées sur des adaptations du PageRank. Il existe également des approches du type apprendre-à-ordonner (\textit{learning-to-rank}) appliquées au web des données (par exemple~\cite{dali2012query}). Ces techniques dépendent de la disponibilité des jugements de pertinence pour l'apprentissage (bien que des mesures indirectes de quantités corrélées puissent parfois être utilisées, par exemple le nombre de fois qu'un agent a accédé à la représentation d'une entité).

Avec l'algorithme LDRANK, nous adoptons une stratégie fondée sur la combinaison consensuelle de plusieurs sources de connaissances afin de modifier un algorithme de type PageRank. Chaque source de connaissance permet de construire une distribution de probabilité sur les entités. Une telle distribution représente l'importance \textit{a priori} de chaque entité. Dans ce contexte, il faut entendre par connaissance \textit{a priori}, toute forme de connaissance contextuelle qui ne dépend pas de la connaissance de la structure du graphe matérialisé par l'ensemble de triplets RDF qui relient explicitement les entités entre elles. La prise en compte de cette structure de graphe se fait à travers la convergence du processus de marche aléatoire du PageRank. Mais ce dernier processus est biaisé par la combinaison convexe de la matrice stochastique correspondant à la structure explicite du graphe RDF, avec une matrice de rang~1 correspondant à une combinaison consensuelle des distributions obtenues à partir des différentes sources de connaissance contextuelle à disposition. L'une de ces sources de connaissance \textit{a priori} est obtenue à partir d'une analyse sémantique latente itérée des données textuelles associées aux entités. Cette dernière stratégie, ainsi que la proposition de combiner consensuellement plusieurs sources de connaissances \textit{a priori} dans le contexte d'un algorithme de type PageRank nous semblent former des contributions nouvelles par rapport à l'état de l'art. Dans cet article, nous montrons que cette approche produit des ordonnancements d'une qualité significativement meilleure à celle des ordonnancements produits par les stratégies de l'état de l'art basées sur des modifications du PageRank.

Finalement, nous montrons également le potentiel de LDRANK en l'appliquant à un contexte de construction de snippets sémantiques. Un snippet est un extrait d'une page web calculé au moment de l'analyse de la requête et devant aider l'utilisateur à décider de la pertinence de la page web par rapport à son besoin d'information. Un snippet sémantique vise à améliorer ce processus de décision et d'exploration en rendant explicites les relations entre un besoin d'information et les entités les plus pertinentes présentes dans une page web. Pour ce faire, nous utilisons d'abord l'algorithme LDRANK pour détecter au sein d'une page web les entités les plus susceptibles d'aider l'utilisateur à répondre à son besoin d'information. Puis, nous adoptons une approche par apprentissage automatique pour associer à chaque entité élue par LDRANK les passages de la page web qui permettent le mieux de mettre en évidence la relation entre l'entité et le besoin d'information.

Dans la section~\ref{related-works}, nous introduisons dans un premier temps les travaux relatifs à l'amélioration des snippets pour le web des documents et pour le web des données, puis nous mentionnons également les différentes stratégies existantes pour l'ordonnancement d'entités d'un graphe RDF. Dans la section~\ref{dataset}, nous décrivons la construction par crowdsourcing d'un jeu de données pour l'évaluation de l'ordonnancement guidé par une requête d'entités LOD. Dans la section~\ref{ldrank}, nous présentons l'algorithme LDRANK et son évaluation comparative avec les approches de l'état de l'art. Dans la section~\ref{ensen}, nous introduisons ENsEN (Enhanced Search Engine), le système logiciel que nous avons développé pour générer des snippets sémantiques. Dans la section~\ref{ml}, nous décrivons une approche par apprentissage automatique au cœur de ENsEN et permettant d'associer aux entités du LOD les plus importantes (au sens du résultat de l'exécution de LDRANK) les passages de la page web les plus à même d'illustrer la relation sémantique entre cette entité et le besoin d'information de l'utilisateur. Dans la section~\ref{enseneval}, nous présentons les résultats d'une évaluation par crowdsourcing du système ENsEN, avant de conclure en section~\ref{conclusion}.

\section{Travaux connexes}\label{related-works}

Nous mentionnons tout d'abord des travaux qui génèrent des snippets à partir de documents RDF natifs. \citeauthor{ge2012incorporating} et \citeauthor{penin2008snippet} s'intéressent à la génération de snippets pour la recherche d'ontologies. \citeauthor{bai2008rdf} génèrent des snippets pour un moteur de recherche sémantique.

Dans~\cite{penin2008snippet}, les auteurs commencent par identifier un sujet thématique grâce à un algorithme de clustering hiérarchique hors-ligne. Ensuite, ils calculent une liste de triplets RDF (c.à.d. des ensembles de triplets RDF connectés) sémantiquement proches du thème. Enfin, grâce à une mesure de similarité fondée sur Wordnet, ils classent les triplets RDF sélectionnés en considérant les propriétés structurelles du graphe RDF et les caractéristiques lexicales des termes présents dans l'ontologie.

Dans~\cite{ge2012incorporating}, les auteurs commencent par transformer le graphe RDF en un graphe qui met en relation des paires de termes et pour lequel chaque arête est associée à un ensemble de triplets RDF. Leur objectif est de construire une représentation compacte des relations qui existent entre les termes de la requête. Ces relations sont à trouver dans le graphe RDF. Pour ce faire, les auteurs décomposent le graphe d'association des termes en composantes dont le rayon ne doit pas dépasser une valeur fixée (c'est un paramètre de l'algorithme) afin d'éviter la découverte de relations trop lointaines entre termes de la requête. Il s'agit ensuite de chercher au sein de ces composantes des sous-graphes connexes qui relient des termes de la requête. Le snippet est alors une somme de ces sous-graphes connexes.

Dans~\cite{bai2008rdf}, les auteurs commencent par attribuer un thème au document RDF. Pour cela, ils utilisent un prédicat tel que \texttt{p:primaryTopic} s'il existe, sinon ils s'appuient sur une heuristique fondée sur la comparaison des URIs des entités candidates pour représenter le thème avec l'URL du document RDF. Ensuite, ils ont proposé un algorithme pour l'ordonnancement des triplets RDF. A ce propos, il semble intéressant de noter comment ils utilisent les propriétés~: pour chaque propriété, ils définissent son importance par rapport aux autres propriétés d'un schéma donné, ils introduisent également les notions de propriétés corrélatives (par exemple \texttt{foaf:name} et \texttt{foaf:family}) et exclusives (par exemple \texttt{foaf:name} et \texttt{foaf:surname}). Enfin, ils utilisent cet algorithme d'ordonnancement pour présenter à l'utilisateur un ensemble de relations entre les triplets proches de la requête et les triplets proches du thème du document RDF.

Pour résumer, comme \citeauthor{ge2012incorporating} nous pensons que posséder des données structurées issues d'un graphe RDF offre la possibilité de trouver des relations non triviales entre les termes de la requête eux-mêmes, mais aussi entre les termes de la requête et les concepts les plus importants du document. En outre, nous suivons également \citeauthor{penin2008snippet} à propos de la nécessité de concevoir un algorithme d'ordonnancement de triplets RDF qui tienne compte à la fois de la structure du graphe et des propriétés lexicales des données textuelles qui peuvent être associées aux n{\oe}uds ou aux arêtes du graphe.

Les travaux précédents exploitent des documents RDF natifs, mais en général les entités du LOD peuvent provenir (i) soit d'un jeu de données du LOD (elles peuvent alors être rassemblées via par exemple des requêtes SPARQL), (ii) soit des annotations sémantiques intégrées à une page web (par exemple, en utilisant RDFa, des microdonnées, ou bien des microformats), ou bien (iii) de la détection automatique des entités dans le texte d'une page web (au moyen par exemple de DBpedia Spotlight).  Or, parmi les approches qui permettent d'améliorer les snippets pour le web des documents en utilisant le web des données \cite{haas2011enhanced}\cite{steiner2010google}, aucune ne repose sur la détection automatique d'entités~: seules les annotations encapsulées explicitement sont utilisées. \citeauthor{haas2011enhanced} utilisent des métadonnées structurées (c.à.d.  des données encodées grâce au formalisme RDFa, ou par le biais de microformats) ainsi que plusieurs techniques d'extraction ad-hoc d'information  pour améliorer les snippets avec des éléments multimédias, des paires clé~/~valeur et des fonctionnalités interactives. Ainsi, en combinant les métadonnées créées par les éditeurs des documents avec des données structurées obtenues par des extracteurs ad-hoc conçus spécialement pour quelques sites populaires, les auteurs de ce travail sont capables de construire des snippets enrichis pour de nombreux résultats d'une requête. Ils ont choisi explicitement de ne pas utiliser directement le LOD afin d'éviter le problème du transfert de confiance entre le web des documents et le web des données. En effet, ils soutiennent que la qualité des processus éditoriaux qui génèrent des parties du web des données à partir du web des documents (par exemple la transformation de Wikipedia en DBpedia) ne peut pas être contrôlée. Par conséquent, de leur point de vue, utiliser le LOD à travers un processus de détection automatique d'entités pour enrichir des snippets s'accompagnerait du risque jugé trop important d'introduire du bruit incontrôlé dans les résultats. De plus, Google Rich Snippet~\cite{steiner2010google} est une initiative similaire qui s'appuie exclusivement sur les métadonnées structurées rédigées par les éditeurs des pages web.

Enfin, une étude faite en 2012~\cite{bizer2013deployment} sur plus de 40 millions de sites web du Corpus Common Crawl montre que seul 5,64\,\% des sites intègrent des données structurées. Cependant, près de 50\,\% des 10~000 premiers sites de la liste Alexa des sites web les plus populaires possédaient des données structurées. En outre, les auteurs de l'étude affirment que (nous traduisons)~: ``Les sujets des données [structurées] [\ldots] semblent être en majeure partie déterminés par les principaux consommateurs qui constituent la cible de ces données~: Google, Yahoo!, Bing''. Ainsi, il nous semble qu'il y a aujourd'hui un besoin évident pour un algorithme qui permette d'exploiter les données structurées issues d'un processus de détection automatique d'entités dans une page web avec une confiance suffisante dans leur qualité, et ce de manière à permettre l'émergence d'applications qui utilisent conjointement le web des données et le web des documents. Dans ce travail, nous proposons une telle solution à travers un algorithme d'ordonnancement des entités d'un graphe du web des données. De plus, nous montrons l'utilité de cet algorithme en l'appliquant au contexte de la construction de snippets sémantiques.

La majorité des approches existantes pour ordonner les entités d'un graphe RDF provenant du web des données sont fondées sur une modification de l'algorithme PageRank. Ainsi, OntologyRank~\cite{ding2004swoogle} (utilisé par Swoogle) modifie la matrice de téléportation pour prendre en compte les types de relations qui existent entre différentes ontologies. D'une manière similaire, PopRank~\cite{nie2005object} propose une version modifiée du PageRank qui tient compte des différents types de prédicats entre entités. Enfin, RareRank~\cite{wei2011rational} modifie la matrice de téléportation pour modéliser l'influence de la proximité thématique entre entités, proximité évaluée grâce à l'introduction d'une mesure de similarité sémantique dont le calcul dépend d'ontologies disponibles pour le domaine considéré.

\section{Construction par crowdsourcing d'un jeu de données pour l'évaluation de l'ordonnancement guidé par une requête d'entités d'un sous-graphe du LOD}\label{dataset}

LDRANK appartient à la classe des algorithmes d'ordonnancement d'entités d'un graphe creux et hétérogène provenant du web des données et accompagné de la représentation d'un besoin d'information sous la forme d'une requête qui est une liste de mots clés. A notre connaissance, il n'existe pas de jeu de données adapté à ce contexte (ce qui peut être vérifié par la lecture d'un état de l'art récent~\cite{roasurvey}). Ainsi, nous avons adopté une approche par crowdsourcing pour construire notre jeu de données qui servira à évaluer notre proposition et à la comparer à l'état de l'art (ce jeu de données est accessible librement\footnote{\url{http://liris.cnrs.fr/drim/projects/ensen/}}). Dans cette section, nous décrivons comment nous avons construit ce jeu de données.

\subsection{Collecte}\label{datasetcollection}

Nous avons sélectionné au hasard 30 requêtes de la collection de requêtes intitulée ``Yahoo! Search Query Tiny Sample'' proposée dans le cadre du projet \textit{Yahoo! Webscope}\footnote{\url{http://webscope.sandbox.yahoo.com/catalog.php?datatype=1}}. Nous avons soumis chaque requête au moteur de recherche Google et nous avons conservé les 5 premières pages web retournées. Pour chacun de ces 150 documents HTML, nous avons extrait le contenu textuel principal en appliquant l'algorithme développé par Kohlschtter, Frankhauser, et Nejdl~\cite{kohlschutter2010boilerplate}. En moyenne, le texte conservé pour chaque page web est fait de 467 mots. Nous avons exécuté DBpedia Spotlight~\cite{mendes2011dbpedia} sur ces textes afin de détecter des entités. En moyenne, 81 entités ont été détectées pour chaque page web. 

Pour chaque document, à partir de ses entités associées, nous construisons un graphe RDF en émettant des requêtes SPARQL sur le jeu de données de DBpedia afin de découvrir tous les triplets RDF qui lient ces entités. A chaque entité du graphe, nous associons un texte obtenu par concaténation de son résumé au sens de DBpedia (c'est-à-dire, le nœud littéral destination du prédicat ``abstract''), et d'une fenêtre de texte (300 caractères) centrée sur le (ou les) passages de la page web où l'entité a été détectée par DBpedia Spotlight. Nous avons supprimé les mots vides et appliqué une étape de racinisation à ce texte. 

\subsection{Génération de micro-tâches}

Étant donnée la longueur de chacun des 150 documents, la tâche qui consisterait à évaluer la pertinence de toutes les entités associées à une page web serait trop lourde pour pouvoir être présentée comme un travail atomique dans le cadre d'une approche par crowdsourcing. Ainsi, il nous faut diviser cette tâche en un ensemble de \textit{micro-tâches}. Dans notre cas, une micro-tâche consiste à évaluer la pertinence des entités annotées dans une seule phrase. Nous découpons le texte d'une page web en phrases en appliquant l'algorithme ICU BreakIterator\footnote{\url{http://icu-project.org/apiref/icu4c/classicu_1_1BreakIterator.html}}. Il y a en moyenne 22 phrases par document. De plus, si une phrase contient plus de 10 entités annotées, le travail est réparti sur plusieurs micro-tâches.

Nous avons utilisé la plateforme de crowdsourcing CrowdFlower\footnote{http://www.crowdflower.com/}. Cette plateforme répartit le travail auprès de plus de 5~millions de contributeurs répartis dans 154 pays, tout en tenant à jour des métriques de qualité pour chaque contributeur. La conception d'une micro-tâche se fait en CML, un idiome XML fourni par CrowdFlower. Pour chaque micro-tâche, nous proposons au contributeur des instructions sur comment réaliser le travail (nous avons essayé de nombreuses approches jusqu'à trouver une formulation qui soit comprise par tous les contributeurs). Nous fournissons au contributeur un thème formé d'un titre (qui est en fait la requête qui a permis d'obtenir le document dont la phrase sur laquelle il travaille est extraite) et d'un court texte (la phrase pour laquelle il doit juger de la pertinence des entités annotées, accompagnée de la phrase précédente et de la phrase suivante dans la page web qui permettent de mieux fixer le contexte). Pour chaque entité annotée dans la phrase, le contributeur doit évaluer la correction et la pertinence de l'annotation. Pour ce faire, nous avons utilisé l'échelle ordinale introduite par Järvelin et Kekäläinen lorsqu'ils proposèrent la métrique DCG (Discounted Cumulative Gain)~\cite{jarvelin2000ir}~: ``irrelevant'' (0), ``marginally relevant'' (1), ``fairly relevant'' (2), et ``highly relevant'' (3). Par ailleurs, chaque question est associée à un court texte qui décrit l'entité considérée (\textit{viz.} le début de son résumé au sens du prédicat ``abstract'' de DBpedia). Nous avons proposé chaque micro-tâche à 10 contributeurs. Ainsi, pour chaque micro-tâche nous avons 10 jugements. Chaque micro-tâche était rémunérée \$.01.

\subsection{Contrôle qualité}

Nous n'avons accepté que des contributeurs qui avaient déjà complété plus de 100 micro-tâches, et dont la précision des réponses, telle que mesurée par la plateforme CrowdFlower, était élevée. Pour évaluer la précision d'un contributeur, CrowdFlower utilise la notion de question-test. Contrairement à une question simple, une question-test est accompagnée de la réponse considérée comme vraie par le concepteur de la question. Ainsi, lors de la conception d'une tâche, l'introduction d'une question-test est considérée comme une bonne pratique.

Nous avons donné au maximum 30 minutes à un contributeur pour répondre à une question. Par ailleurs, nous avons imposé qu'un contributeur passe au minimum 10 secondes sur une tâche avant de fournir une réponse.

Nous avons mesuré l'accord entre contributeurs grâce au coefficient alpha de Krippendorff~\cite{krippendorff2012content}. Ce coefficient utilise par défaut une distance binaire pour comparer deux réponses, mais il est conçu pour pouvoir utiliser également d'autres distances. Pour tenir compte du fait que nous avons utilisé une échelle ordinale qui modélise à la fois la correction et la pertinence, nous avons utilisé la distance symétrique décrite par le tableau~\ref{tab:distance}.

\begin{table}[tab:distance]{Distance symétrique pour le calcul de l'alpha de Krippendorff}
$\begin{array}{c|cccc}
d & 0 & 1 & 2 & 3\\
\hline
0 & 0,00 & 0,50 & 0,75 & 1,00\\
1 & 0,50 & 0,00 & 0,25 & 0,50\\
2 & 0,75 & 0,25 & 0,00 & 0,25\\
3 & 1,00 & 0,50 & 0,25 & 0,00
\end{array}$
\end{table}

Avec ces paramètres, nous avons obtenu un alpha de $0,22$. Selon l'échelle interprétative de Landis et Koch~\cite{landis1977measurement} ce score peut être considéré comme un accord acceptable (l'échelle a été conçue pour le kappa de Fleiss, mais l'alpha de Krippendorff est en tout point compatible avec le kappa). Cependant, par comparaison avec des travaux existants qui ont appliqué du crowdsourcing dans un contexte de recherche d'information, nous pourrions espérer un alpha plus élevé. Par exemple, Jeong \textit{et al.}~\cite{jeong2013crowd} ont obtenu un kappa de Fleiss de $0,41$ (c'est-à-dire, un accord modéré) pour un moteur de recherche qui fait appel à la sagesse des foules pour répondre à certaine requêtes. Cependant, Alonso, Marshall et Najork~\cite{alonsocrowdsourcing} ont obtenu un alpha de Krippendorff qui varie entre $0,03$ et $0,19$ dans le cadre d'une tâche plus subjective~: décider de l'intérêt d'un tweet. Ainsi, afin d'améliorer la qualité de notre jeu de données, nous avons choisi d'isoler les contributeurs qui étaient souvent en désaccord avec la majorité. En retirant les contributeurs qui sont en désaccord avec la majorité dans plus de $41,2\,\%$ de leurs jugements, nous obtenons un alpha de $0,46$. Après cette opération, $96,5\,\%$ des tâches possèdent au moins $3$ jugements, $66\,\%$ des tâches possèdent au moins $5$ jugements, et seulement $0,7\,\%$ des tâches ne possèdent que $1$ jugement.

\subsection{Agrégation des résultats}

Nous avons utilisé le vote à la majorité pour l'agrégation des résultats au sein d'une phrase. Nous avons testé deux approches différentes pour séparer les \textit{ex aequo}~: (i) le maximum de la moyenne de la précision des contributeurs (la précision d'un contributeur est une métrique proposée par la plateforme CrowdFlower et que nous avons mentionnée plus haut), (ii) la valeur la plus élevée. Nous avons pu constater \textit{a posteriori} que ces deux choix donnaient des résultats très similaires lorsque le jeu de données est utilisé pour comparer des algorithmes d'ordonnancement d'entités étant donné un besoin d'information exprimé par une liste de mots clés. Nous avons utilisé la même stratégie de vote à la majorité pour agréger les résultats au niveau d'une page web.

Dans la prochaine section, nous introduisons LDRANK, un algorithme pour l'ordonnancement guidé par une requête d'entités du LOD. Le jeu de données dont nous venons de décrire la construction sera utilisé à la section~\ref{ldrankeval} pour évaluer LDRANK et le comparer à l'état de l'art.

\section{LDRANK, un algorithme pour l'ordonnancement guidé par une requête d'entités du LOD}\label{ldrank}

\subsection{Contexte}

Étant donné la connaissance d'un besoin d'information exprimé sous la forme d'une requête faite d'un ensemble de mots clés, l'algorithme LDRANK fournit un ordonnancement des entités d'un graphe RDF pour lequel des données textuelles sont associées aux nœuds. Pour calculer cet ordonnancement, LDRANK utilise à la fois la structure explicite du graphe et les relations implicites découvertes par analyse du texte associé aux entités. Grâce à cette stratégie double, il est particulièrement adapté aux graphes creux et bruités (c'est-à-dire, avec une proportion importante de nœuds non pertinents selon le besoin d'information exprimé par la requête). De tels graphes apparaissent en particulier suite à l'annotation automatique d'une page web (par exemple, via DBpedia, AlchemyAPI,...). C'est pourquoi nous avons construit le jeu de données introduit à la section précédente de cet article (voir en particulier la sous-section~\ref{datasetcollection}).

LDRANK prend en compte la structure explicite du graphe à travers un algorithme de type PageRank. De plus, il utilise les relations implicites qui peuvent être inférées à partir du texte associé aux entités grâce à une variante originale de l'algorithme de décomposition en valeurs singulières (SVD). Enfin, LDRANK prend également en compte l'ordonnancement des pages web au sein desquelles les entités ont été détectées grâce à une fonction de score d'abord introduite par \citeauthor{fafalios2014post}.

Plus précisément, l'analyse textuelle fondée, sur une application itérée de la décomposition en valeurs singulières, ainsi que l'exploitation de l'ordonnancement fourni par la page de résultats d'un moteur de recherche du web, permettent chacune de construire un vecteur de probabilités exprimant une connaissance a priori (ou opinion) sur l'importance relative des entités (voir les sections~\ref{ldrankserp} et~\ref{ldranksvd}). Ensuite, ces vecteurs de probabilités sont rassemblés grâce à une stratégie d'agrégation consensuelle linéaire d'opinions d'abord introduite par \citeauthor{carvalho2013consensual} (voir section~\ref{ldrankaggreg}). Enfin, nous utilisons ces différentes sources de connaissance a priori, agrégées en une unique distribution de probabilités sur les entités, pour influencer le processus de convergence d'un algorithme de type PageRank vers une distribution de probabilité stable (au sens d'un vecteur propre de valeur propre unité pour un processus de Markov) qui correspond à l'ordonnancement final des entités (voir section~\ref{ldrankpagerank}).

\subsection{Connaissance a priori déduite de l'ordonnancement fourni par la page de résultats d'un moteur de recherche du web}\label{ldrankserp}

\algbegin Algorithme H (Hit Score). Cet algorithme calcule un vecteur de probabilités (\textit{hitdistrib}) qui représente une connaissance \textit{a priori} sur l'importance des entités. Cette connaissance est déduite de l'ordonnancement des pages web au sein desquelles ces entités ont été détectées. L'ordonnancement en question est celui retourné par un moteur de recherche du web à partir du besoin d'information de l'utilisateur exprimé sous la forme d'un ensemble de mots clés. Cette stratégie a été introduite pour la première fois par \citeauthor{fafalios2014post}.
\algstep H1. $A \leftarrow $ la liste des meilleures pages web telles qu'ordonnées par un moteur de recherche du web en réponse à une requête formée d'un ensemble de mots clés.
\algstep H2. $E \leftarrow $ l'ensemble des entités détectées dans ces pages web (par exemple, grâce à l'application de DBpedia Spotlight).
\algstep H3. $docs(e) \equiv $ les documents de $A$ au sein desquels l'entité $e$ a été détectée.
\algstep H4. $rank(a) \equiv $ le rang du document $a$ dans $A$.
\algstep H5. $hitscore(e) \equiv \sum_{a \in docs(e)} (size(A) + 1) - rank(a)$
\algstep H6. $hitdistrib[e] \leftarrow hitscore(e) / \sum_{e' \in E} hitscore(e')$
\algstep H7. [{\it End.}] \quad\slug

\subsection{Connaissance a priori déduite d'une analyse sémantique latente itérée des données textuelles associées aux entités}\label{ldranksvd}

\aalgbegin Algorithme S (SVD itératif). Cet algorithme calcule un vecteur de probabilités (\textit{svddistrib}) qui représente une connaissance \textit{a priori} sur l'importance des entités. Cette connaissance est fondée sur une analyse du texte associé à chaque entité.
\algstep S1. [{\it Matrice initiale.}] $R \leftarrow $ la matrice creuse entité-terme (c'est-à-dire, avec les entités en ligne, et les termes en colonnes) au format CCS (Compressed Column Storage)\footnote{\url{http://netlib.org/linalg/html_templates/node92.html}}.
\algstep S2. [{\it Initialisation des entités importantes.}] $info\_need \leftarrow $ un ensemble d'entités qui est formé de l'union des entités détectées dans le texte de la requête et de l'entité qui possède le meilleur hitscore (la présence de cette dernière entité est nécessaire pour les cas où aucune entité ne serait détectée à partir des mots clés de la requête). Nous supposons que ces entités sont probablement proches du besoin d'information de l'utilisateur.
\algstep S3. [{\it Premier SVD.}] $(U,S,V^T) \leftarrow svdLAS2A(R, nb\_dim)$ Calcule la décomposition en valeurs singulières (SVD) de $R$ au rang $k = nb\_dim$. Puisque $R$ est très creuse, nous utilisons l'algorithme \textit{las2} développé par Michael W.  Berry~\cite{berry1992large} afin de calculer la décomposition~: $R_k = U_k S_k V_k^T$ avec $U_k$ et $V_k$ des matrices orthogonales, $S_k$ une matrice diagonale, telles que $\|R-R_k\|_F$ soit minimisée (c'est-à-dire que du point de vue de la norme de Frobenius, $R_k$ est la meilleure approximation au rang $k$ de $R$).
\algstep S4. [{\it Coordonnées des entités dans l'espace réduit.}] $SUT \leftarrow S U^T$ Dans le nouvel espace à $k$ dimensions, cette opération met à l'échelle les coordonnées des entités (c'est-à-dire, les lignes de $U$) grâce aux facteurs de contraction/dilatation correspondants de $S$. Ainsi, nous obtenons les coordonnées des entités dans le nouvel espace réduit (c'est-à-dire, les colonnes de $SUT$).
\algstep S5. $prev\_norms \leftarrow $ les normes euclidiennes des entités dans l'espace réduit.
\algstep S6. [{\it Matrice mise à jour.}] $R' \leftarrow $ $R$ où les lignes correspondant aux entités de l'ensemble $info\_need$ ont été multipliées par le paramètre $stress$ (puisque $R$ est au format CCS, il est plus pratique de faire cette opération sur la transposée de $R$).
\algstep S7. [{\it Second SVD.}] $(U',S',V'^T) \leftarrow svdLAS2A(R', nb\_dim)$
\algstep S8. [{\it Mise à jour des coordonnées des entités dans l'espace réduit.}] $SUT' \leftarrow S' U'^T$
\algstep S9. $norms \leftarrow $ mise à jour des normes euclidiennes des entités dans l'espace réduit.
\algstep S10. [{\it Éloignement des entités par rapport à l'origine de l'espace réduit.}.] \\$svdscore(e) \equiv norms[e] - prev\_norms[e]$.
\algstep S11. $svddistrib[e] \leftarrow svdscore(e) / \sum_{e'} svdscore(e')$
\algstep S12. [{\it End.}] \quad\slug

Nous pouvons maintenant introduire la propriété de la décomposition en valeurs singulières sur laquelle repose l'algorithme~S. Pour une forte réduction dimensionnelle (c.à.d. pour une faible valeur de $k$), la transformation $S_k U^T$ tend à placer les entités qui étaient orthogonales à de nombreuses entités dans l'espace des lignes de $R$ proches de l'origine de l'espace $k$-dimensionnel résultant. En effet, comme vu plus haut, le SVD peut être vu comme un algorithme d'optimisation, or pour minimiser l'erreur due à l'impossibilité pour une entité d'être orthogonale à plus de $k$ entités non colinéaires, cette entité doit être placée aussi proche que possible de l'origine de l'espace réduit. Un argument similaire peut être utilisé pour montrer qu'une entité colinéaire à de nombreuses entités dans l'espace des lignes de $R$ aura aussi tendance à se trouver proche de l'origine de l'espace réduit $k$-dimensionnel. Ainsi, et pour résumé, les entités éloignées de l'origine dans l'espace réduit ont tendance à posséder des relations ``intéressantes'' avec d'autres entités éloignées de l'origine. Par ailleurs, les entités qui sont dans la direction de plus grande variation des données dans l'espace initial sont les mieux alignées avec l'axe correspondant à la plus grande valeur singulière dans l'espace réduit (c.à.d. l'axe qui a subi l'extension la plus forte), elles ont donc tendance à être les plus éloignées de l'origine de l'espace réduit. Ainsi, le principe à l'{\oe}uvre dans l'algorithme~S consiste à augmenter artificiellement l'importance des entités sémantiquement proches de la requête afin de les forcer dans la direction de plus grande variation des données, et à observer parmi les autres entités celles qui, suite à cette opération, s'éloignent le plus de l'origine de l'espace réduit. Grâce à l'argument énoncé ci-dessus, ces dernières entités peuvent ainsi être qualifiées de potentiellement proches du besoin d'information.

Nous avons obtenu les meilleurs résultats expérimentaux avec une réduction sur un espace uni-dimensionnel (c'est-à-dire avec $nb\_dim = 1$ aux étapes~S3 et~S7 de l'algorithme~S), et avec un facteur de stress de $1000.0$ (voir l'étape~S6 de l'algorithme~S).

\subsection{Stratégie pour l'agrégation de plusieurs sources de connaissance a priori}\label{ldrankaggreg}

Nous considérons \textit{hitdistrib} (voir algorithme~H), \textit{svddistrib} (voir algorithme~S), et la distribution équiprobable (\textit{equidistrib}) comme trois opinions expertes (ou trois sources de connaissance \textit{a priori}) sur l'importance des entités. Afin d'agréger ces opinions, nous appliquons l'algorithme de Carvalho et Larson~\cite{carvalho2013consensual} qui permet de déterminer itérativement une combinaison linéaire optimale de plusieurs vecteurs de probabilités. A chaque pas de cet algorithme itératif, l'expert \textit{i} re-calcule la distribution représentant son opinion sous la forme d'une combinaison linéaire des distributions de tous les experts. Dans cette combinaison linéaire, le poids affecté par l'expert \textit{i} à la distribution de l'expert \textit{j} est proportionnel à la distance séparant ces deux distributions. Les auteurs de ce travail ont défini une distance telle que ce processus itératif converge toujours vers une unique distribution dite consensuelle. Nous appellerons le vecteur de probabilités consensuel résultant~: \textit{finaldistrib}.

\subsection{LDRANK}\label{ldrankpagerank}

L'algorithme PageRank~\cite{page1999pagerank} transforme la matrice d'adjacence ($M$) d'un réseau de pages web en une matrice $H$ qui est \textit{stochastique} (c'est-à-dire que la somme de chaque ligne de $H$ vaut $1$) et \textit{primitive} (c'est-à-dire qu'il existe un entier $k$ tel que $H^k > 0$), assurant ainsi l'existence d'un vecteur stationnaire (c'est-à-dire, le vecteur propre positif correspondant à la valeur propre $1$). Ce vecteur stationnaire est un vecteur de probabilités qui peut être interprété comme représentant \textit{l'importance} de chaque page web en modélisant rigoureusement la proposition intuitive selon laquelle une page web importante est référencée par des pages web importantes. De plus, cette distribution stationnaire peut être calculée de manière efficace grâce à la méthode dite de la puissance itérée tenant compte du fait que la matrice stochastique est creuse.

Dans la version originale de l'algorithme PageRank, aucune hypothèse n'est faite sur la probabilité \textit{a priori} de l'importance d'une page web avant que ne soit entamée l'analyse de la structure explicite du graphe. En d'autres termes, $M$ est d'abord transformée en une matrice stochastique $S$ en remplaçant chaque ligne vide par la distribution équiprobable (\textit{equidistrib})~; puis $S$ est transformée en une matrice primitive $H$ par combinaison convexe avec une matrice de rang $1$ dite de téléportation ($T$)~: $H = \alpha S + (1-\alpha)T$ où chaque ligne de $T$ est la distribution équiprobable (\textit{equidistrib}).

Pour l'algorithme LDRANK, au lieu d'utiliser la distribution équiprobable, nous utilisons la distribution consensuelle (\textit{finaldistrib}) introduite plus haut (voir section~\ref{ldrankaggreg}). Nous avons obtenu les meilleurs résultats expérimentaux pour $0.6 \leq \alpha \leq 0.8$. De plus, nous avons fixé à $1E-10$ la valeur de la précision de convergence qui contrôle la terminaison de la méthode de la puissance itérée utilisée pour calculer le vecteur stationnaire.

Enfin, LDRANK est disponible librement en ligne\footnote{\url{http://liris.cnrs.fr/drim/projects/ensen/}}.

\subsection{Évaluation de LDRANK}\label{ldrankeval}

Nous comparons quatre stratégies d'ordonnancement. Chacune utilise une source de connaissance \textit{a priori} différente pour informer l'algorithme PageRank~: la version classique pour laquelle la connaissance \textit{a priori} sur l'importance des entités est nulle et correspond donc  à la distribution équiprobable (nous appelons cette stratégie EQUI)~; la version modifiée avec l'utilisation du hitscore comme source de connaissance \textit{a priori} (cette version est due à \citeauthor{fafalios2014post} voir section~\ref{ldrankserp}, nous nommons cette stratégie HIT)~; la version modifiée avec notre nouvelle source de connaissance \textit{a priori} basée sur une utilisation itérée de la décomposition en valeurs singulières (voir section~\ref{ldranksvd}, nous nommons cette stratégie SVD)~; et LDRANK qui utilise la combinaison consensuelle de trois sources différentes de connaissance \textit{a priori}.

Afin de comparer les quatre stratégies (EQUI, HIT, SVD et LDRANK), nous utilisons la métrique NDCG (Normalized Discounted Cumulative Gain). Le DCG (Discounted Cumulative Gain) au rang $r$ est défini ainsi~: $DCG_r = rel_1 + \sum_{i=1}^r \frac{rel_i}{log_2 i}$ (avec $rel_i$ représentant la pertinence estimée du résultat au rang $i$, la pertinence est représentée sur une échelle discrète qui contient généralement bien moins de valeurs que le nombre total de résultats à ordonner, nous utilisons une échelle à quatre gradations, voir section~\ref{dataset}). Le NDCG au rang $r$ est égal au DCG au rang $r$ normalisé par le DCG au rang $r$ d'un ordonnancement optimal.

Les résultats sont présentés sur la figure~\ref{fig:ndcg}. Nous observons que les stratégies SVD et HIT ont des performances comparables. Cependant, elles sont clairement surpassées par leur combinaison consensuelle LDRANK. De plus, comme notre implémentation tient systématiquement compte du caractère creux des matrices, nous obtenons également de bonnes performances en termes de temps d'exécution (voir figure~\ref{fig:perfs}). La stratégie SVD prend plus de temps que la stratégie HIT puisqu'elle nécessite le calcul de la décomposition en valeurs singulières. Le temps supplémentaire pris par la stratégie LDRANK s'explique par le temps de convergence du processus itératif qui calcule la combinaison consensuelle des différentes distributions. Enfin, nous avons mené des expérimentations similaires en considérant toutes les arêtes du graphe bidirectionnelles (en effet, le sens de l'interprétation d'un prédicat RDF est en pratique souvent arbitraire). Les performances relatives des algorithmes sont alors similaires, mais en valeur absolue les scores NDCG sont légèrement meilleurs.

Notons qu'à travers ces expérimentations, en plus d'introduire une nouvelle stratégie d'ordonnancement efficace basée sur une utilisation originale de la réduction dimensionnelle par décomposition en valeurs singulières, nous montrons également que plusieurs stratégies basées sur une modification de la matrice de téléportation de l'algorithme PageRank peuvent être avantageusement combinées lorsqu'elles sont envisagées en tant que sources distinctes de connaissance \textit{a priori} sur l'importance des entités du graphe.

\begin{figure}[fig:ndcg]{Comparaison des scores NDCG pour les quatre stratégies (EQUI, HIT, SVD et LDRANK)}
\includegraphics[width=4in, keepaspectratio]{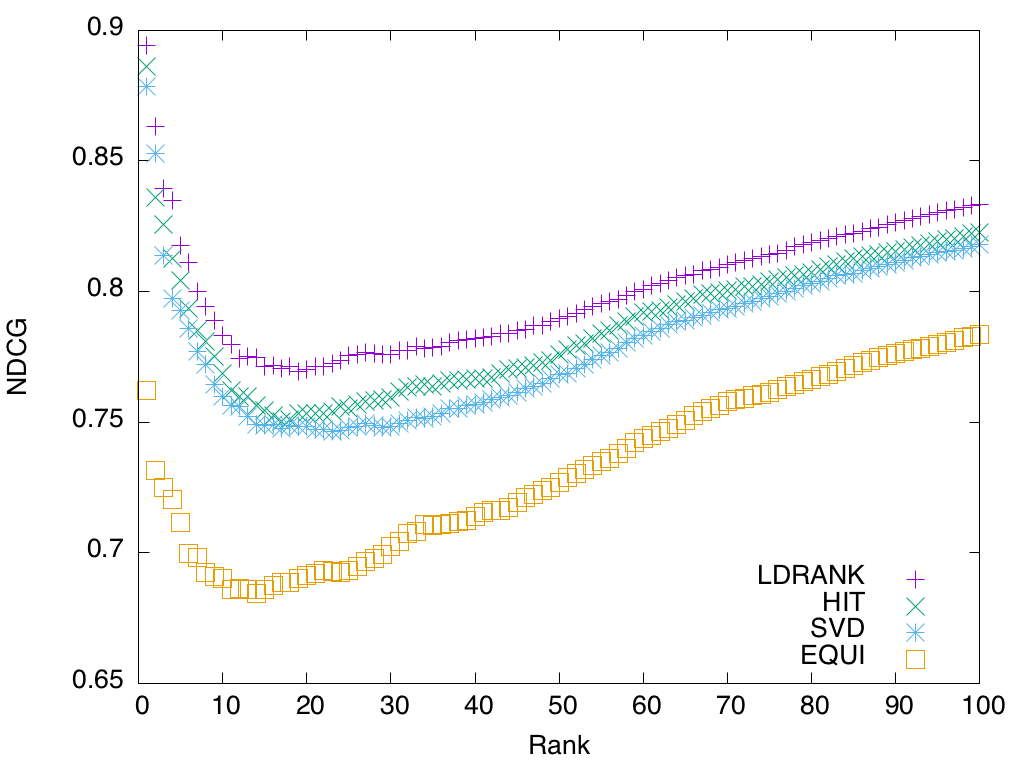}
\end{figure}

\begin{figure}[fig:perfs]{Comparaison des temps d'exécution pour les quatre stratégies (avec un processeur~: 2.9~GHz Intel Core i7, et une mémoire~: 8~GB 1600~MHz DDR3)}
\includegraphics[width=4in, keepaspectratio]{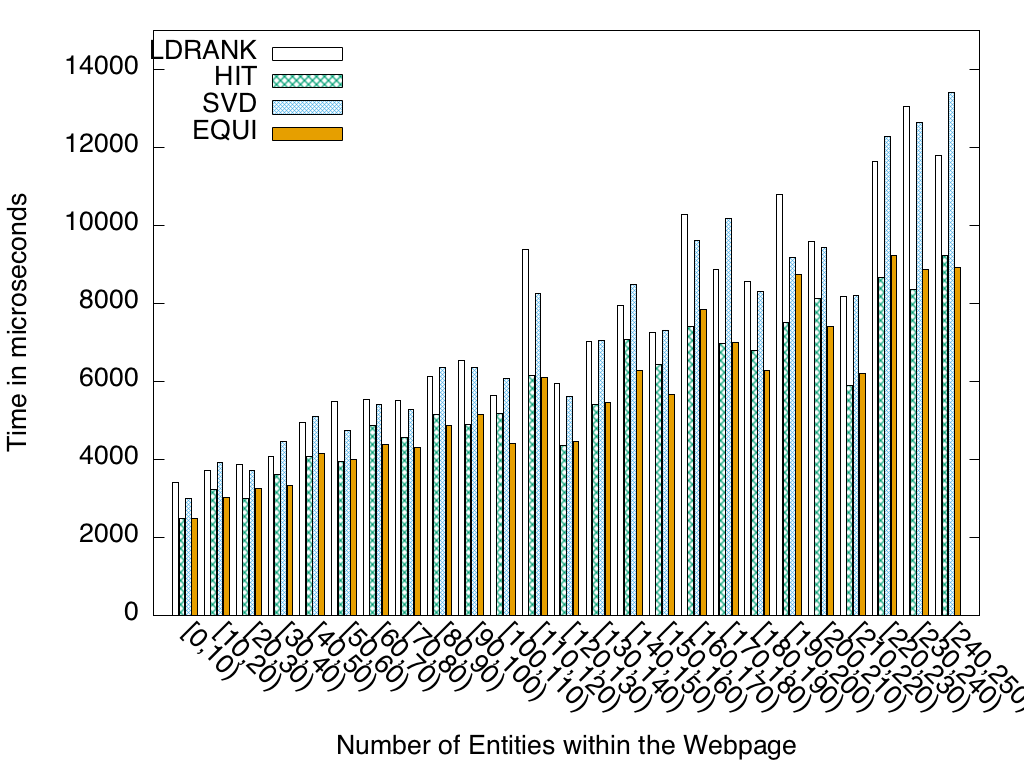}
\end{figure}

\section{Présentation de ENsEN pour la génération de snippets sémantiques}\label{ensen}

\begin{figure}[fig:schema]{Flux des données et traitements pour ENsEN}
\includegraphics[width=4.9in, keepaspectratio]{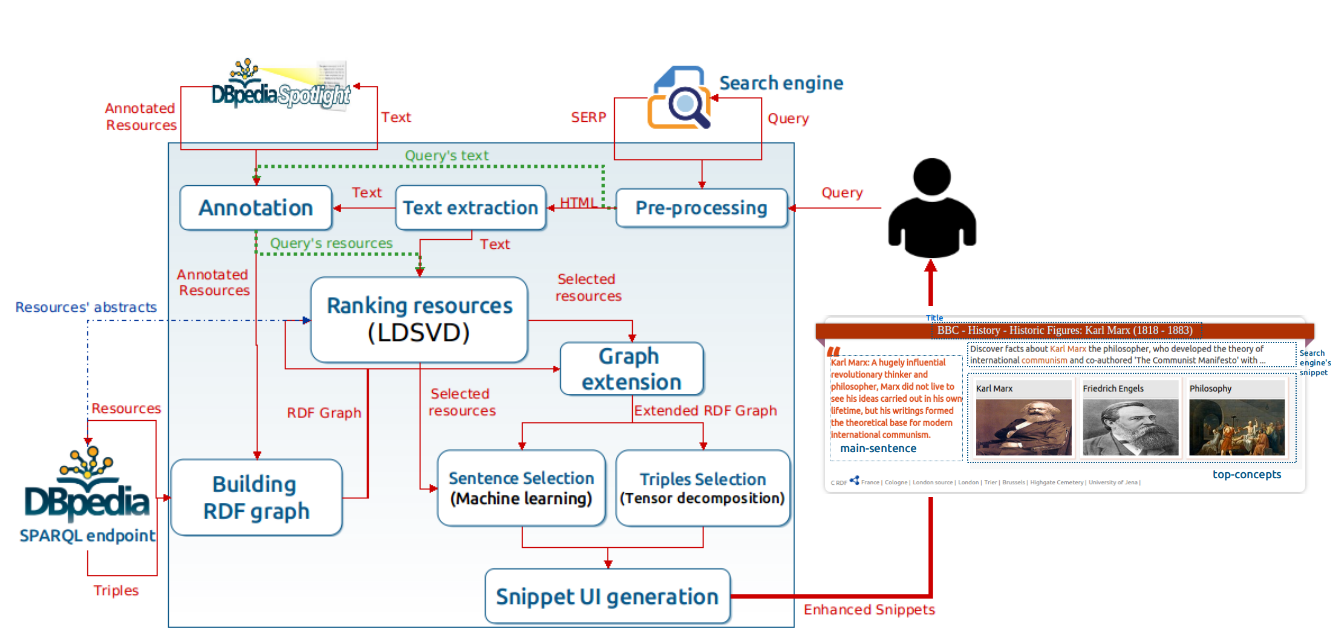}
\end{figure}

Afin de convaincre de l'utilité et de l'efficacité de l'algorithme LDRANK, nous l'utilisons dans le contexte de ENsEN (Enhanced Search Engine), un système que nous avons conçu pour construire des snippets sémantiques (voir figure~\ref{fig:schema}, une démonstration \textit{live} du système est disponible en ligne, voir une précédente note de bas de page pour l'URL). Nous présentons maintenant le flux des données et les traitements qui produisent un snippet sémantique à partir d'une requête, et ce afin de mettre en avant le rôle joué par l'algorithme LDRANK dans ce processus. 

Étant donnée une requête, nous obtenons la page de résultats d'un moteur de recherche du web (nous avons utilisé Google pour nos expériences). Pour chaque page web résultat, nous utilisons le service de détection automatique d'entités DBpedia Spotlight afin d'obtenir un ensemble d'entités. De la même manière, nous trouvons les entités associées aux termes de la requête. A partir de cet ensemble d'entités et par l'intermédiaire de requêtes émises auprès d'un point d'accès SPARQL de DBpedia, nous obtenons un graphe en trouvant toutes les relations qui existent entre ces entités dans le jeu de données DBpedia. 

A chaque entité nous associons un texte obtenu par concaténation de son résumé DBpedia et de fenêtres textuelles centrées sur les occurrences de l'entité dans la page web (nous utilisons des fenêtres de 300 caractères). De plus, nous supprimons les mots vides et appliquons un enracineur\footnote{\url{http://snowball.tartarus.org/}}. Nous exécutons l'algorithme LDRANK avec en entrée ce graphe aux n{\oe}uds décorés de texte ainsi que les entités découvertes dans la requête, et nous récupérons un ordonnancement des entités. Pour chacune des meilleures entités, nous construisons des vignettes affichées sur le snippet. 

A partir d'un point d'accès SPARQL de DBpedia, nous procédons à une extension 1-hop des meilleures entités afin d'augmenter le nombre de triplets parmi lesquels nous cherchons ensuite les plus importants en terme d'une analyse des liens du graphe. Pour ce faire, nous construisons un tenseur d'ordre~3 à partir du graphe étendu~: chaque prédicat correspond à une couche horizontale d'ordre~2 qui représente la matrice d'adjacence pour la restriction du graphe à ce prédicat. Nous calculons ensuite la décomposition PARAFAC du tenseur en une somme de facteurs (qui sont des tenseurs de rang~1 et d'ordre~3), et nous l'interprétons d'une façon similaire à Franz~{\it et al.}~\cite{franz2009triplerank}~: pour chacune des meilleures entités (au sens du LDRANK), nous sélectionnons les facteurs auxquels elle contribue le plus (en tant que sujet ou en tant qu'objet), et pour chacun de ces facteurs, nous sélectionnons les triplets qui ont les prédicats avec les meilleurs scores. Ainsi, nous sommes capables d'associer à chacune des meilleures entités un ensemble de triplets que nous faisons apparaître sur le snippet dans la description de l'entité. 

Enfin, nous avons employé une approche par apprentissage automatique afin de sélectionner de courts passages de la page web qui viennent accompagner la description de chaque entité. Nous décrivons plus avant cette approche dans la prochaine section.

\section{Découverte par apprentissage automatique des passages d'une page web à associer à une entité}\label{ml}

\subsection{Objectifs}

Nous employons une approche par apprentissage automatique supervisé afin d'associer aux meilleurs entités, au sens de notre algorithme LDRANK, les passages de la page web les plus susceptibles de mettre en avant une relation intéressante entre cette entité et le besoin d'information de l'utilisateur. Nous modélisons la découverte de ces associations entre une entité et des passages d'une page web sous la forme d'un problème de classification binaire. C'est-à-dire qu'étant donnée une entité, pour chaque passage de la page web au sein duquel l'entité a été détectée, il faut décider si ce passage illustre bien la relation entre cette entité et le besoin d'information de l'utilisateur.

Nous commencerons, en section~\ref{mlrelated}, par introduire des travaux existants qui répondent à des problèmes similaires. Puis, en section~\ref{mlfeatures}, nous détaillerons les différentes variables que nous avons choisies d'utiliser pour l'apprentissage. En section~\ref{mldataset}, nous expliquerons comment nous avons construit le jeu de données pour l'apprentissage, et en section~\ref{mlunbalanced} nous montrerons comment nous avons ré-équilibré le jeu d'apprentissage pour rendre les classes comparables. En section~\ref{mlresults1}, nous discuterons les résultats obtenus par différents algorithmes d'apprentissage lorsque toutes les variables sont conservées. Ensuite, en section~\ref{mlfeatureselection}, nous introduirons différentes stratégies pour réduire le nombre de variables utilisées pour la classification, et en section~\ref{mlresults2} nous discuterons les nouveaux résultats obtenus. Enfin, en section~\ref{mlsynthese} nous synthétiserons ces différents résultats en soulignant l'importance de LDRANK dans ce processus.

\subsection{Travaux connexes}\label{mlrelated}
Nous mentionnons maintenant quelques travaux qui adressent un problème comparable au nôtre~: la sélection d'extraits d'une page web étant donnée la connaissance d'une requête exprimant le besoin d'information de l'utilisateur. Notre contexte diffère principalement par la présence d'une couche sémantique issue de la détection d'entités dans la page web et de leur ordonnancement par LDRANK. Mais les deux problèmes restent suffisamment proches pour que nous puissions bénéficier d'une étude de l'état de l'art pour ce problème classique de la sélection d'extraits d'une page web étant donnée une requête. Ainsi, nous présentons quelques travaux représentatifs qui adoptent des stratégies fondées sur l'apprentissage automatique.

Dans~\cite{wang2007learning}, les auteurs introduisent des variables qui prennent en compte à la fois le contenu de la page web et également le contenu contextuel qui provient des textes d'ancrage de liens qui pointent vers la page web. Dans ce travail, la granularité de l'extrait est celle de la phrase. Les auteurs modélisent la pertinence par le décompte du nombre de termes de la requête présent dans une phrase. Ils modélisent l'importance par le décompte normalisé des termes qui sont les plus fréquents dans la page et qui apparaissent dans la phrase considérée, ainsi que par le décompte des termes constitutifs du titre de la page , et enfin par le décompte des termes qui apparaissent dans les textes d'ancrage des liens qui pointent vers la page. 

Les auteurs expérimentent avec deux stratégies d'apprentissage dont ils comparent les performances. La première stratégie consiste à modéliser le problème de la sélection de phrases comme un problème de classification binaire, ensuite résolu par l'entraînement d'un classifieur SVM. Pour prendre en compte le déséquilibre entre le nombre d'instances dans chacune des deux classes, les auteurs séparent le paramètre qui règle la pénalité pour les erreurs de classification en deux composantes avec un facteur différent pour chacune des deux classes. La seconde stratégie consiste à adopter une approche du type ``apprendre à ordonner'' en apprenant une fonction d'ordonnancement qui doit placer les phrases sélectionnées avant les phrases non sélectionnées \textit{au sein d'une page donnée du jeu d'apprentissage} (contrairement à la première stratégie de type classification binaire où il s'agissait de distinguer les phrases sélectionnées de celles non sélectionnées sur \textit{l'ensemble des pages du jeu d'apprentissage}). Pour cette seconde stratégie, les auteurs utilisent l'algorithme Ranking~SVM.

Les auteurs de~\cite{wang2007learning} expérimentent sur les données de la Web Track de TREC-2003. Pour chaque requête d'une sélection aléatoire de 10 requêtes, ils ont retenu 10 pages pertinentes, 5 pages non pertinentes parmi les mieux classées (du point de vue de l'algorithme BM25), et 5 pages non pertinentes choisies aléatoirement. Deux évaluateurs humains ont construit manuellement des résumés pour chacune des pages en sélectionnant les meilleures phrases (le plus souvent trois d'entre elles) étant donnée la requête et son contexte (c'est-à-dire la description détaillée présente dans le jeu de données TREC). Pour les deux stratégies présentées ci-dessus, les auteurs utilisent pour le SVM un noyau Gaussien dont le paramètre de l'écart type est déterminé grâce à une approche par validation croisée avec une division en trois échantillons. Les métriques utilisées pour l'évaluation sont la précision, le rappel, et le F1-score. Les auteurs obtiennent les meilleurs résultats en prenant en compte le contexte (à travers les textes d'ancrage) et en utilisant la stratégie de type ``apprendre à ordonner''.

Dans~\cite{metzler2008machine}, les auteurs mènent un travail similaire à celui que nous venons d'introduire (\cite{wang2007learning}), aux différences près qu'il comparent trois type d'algorithmes d'apprentissage adaptés aux approches \textit{apprendre à ordonner} (Ranking~SVM, Support Vector Regression, et Gradient Boosted Decision Trees), qu'ils introduisent des variables indépendantes de la requête (viz., la longueur de la phrase, et sa position dans la page web), et qu'ils procèdent à des évaluation sur un jeu de données bien plus volumineux (viz., TREC Novelty Track 2002, 2003 et 2004). Les meilleurs résultats ont été atteints par l'algorithme GBDT.

Dans~\cite{ageev2013improving}, les auteurs suivent les résultats introduits ci-dessus (\cite{metzler2008machine}) et utilisent un algorithme de Gradient Boosting Regression Tree (GBRT). Cependant, ce travail se distingue par l'introduction de variables supplémentaires de type comportementales (par exemple, le temps passé par le curseur de la souris sur un fragment de texte, le temps durant lequel un fragment de texte était visible sur l'écran, etc.). Pour construire un jeu de données d'apprentissage incluant ces nouvelles variables, les auteurs ont adopté une approche de type \textit{crowdsourcing} à travers la plateforme Amazon MTurk. Chaque participant devait effectuer des tâches qui chacune consistait à répondre à une question. L'expérience était présentée au participant comme un jeu pour lequel il s'agissait de répondre correctement à un maximum de questions en un temps imparti. Par ailleurs, pour permettre l'enregistrement efficace des variables comportementales, les auteurs sélectionnent d'abord les phrases de la page web qui contiennent au moins un des termes de la requête. Puis ils découpent ces phrases grâce à une fenêtre mouvante (dont la taille d'au minimum 3 mots est un paramètre), et ne conservent que les fragments qui contiennent au moins un terme de la requête.

Enfin, dans~\cite{lehmann2012defacto}, les auteurs proposent l'algorithme DeFacto pour valider la véracité d'un triplet RDF du LOD par la découverte de fragments de pages web qui serviront de sources d'information confirmant le fait exprimé par le triplet. Au cœur de ce travail, se trouve à nouveau une approche par apprentissage automatique supervisé. Ce travail se base sur le système BOA issu de travaux antérieurs des mêmes auteurs. BOA permet d'associer à un prédicat RDF un pattern en langue naturelle. Les auteurs commencent par filtrer les phrases de la page web, et ne conservent que celles dans lesquelles apparaissent les labels associés au sujet et à l'objet du prédicat dont il s'agit de vérifier la véracité. Les variables utilisées pour l'apprentissage comprennent entre autres~: la présence d'un pattern BOA dans la phrase, la distance séparant les labels correspondant au sujet et à l'objet du triplet, etc. Par ailleurs, mentionnons qu'une partie de ce travail consiste également en la proposition d'un ensemble de métriques pour quantifier la confiance à accorder à une page web étant donné le triplet RDF qu'il s'agit de valider. Nous n'entrons pas plus dans les détails de cette dernière partie car elle n'est pas liée à notre problème.

Dans la section suivante, nous décrivons les variables utilisées pour l'apprentissage. Nos choix ont été guidés par l'analyse des travaux connexes introduits ci-dessus.

\subsection{Choix des variables}\label{mlfeatures}

La particularité de notre approche tient surtout à l'utilisation des entités du LOD détectées dans une page web. Rappelons en préambule qu'à chaque entité nous associons un label, un texte qui en résume la signification et qui est obtenu à partir de DBpedia, une liste d'entités voisines dans le graphe RDF construit à partir des entités détectées dans la page web, et enfin le score calculé par notre algorithme LDRANK. Nous décrivons maintenant les différentes variables utilisées pour apprendre à sélectionner les phrases d'une page web les plus propices à expliquer la relation qui existe entre une entité donnée et le besoin d'information de l'utilisateur tel qu'exprimé par sa requête. Nous procédons à cette description en distinguant trois groupes de variables~: celles indépendantes de la requête et de l'entité sélectionnée, celles dépendantes de la requête mais indépendantes de l'entité sélectionnée, celles dépendantes de l'entité sélectionnée, et enfin celles dépendantes des entités annotées (et non spécifiquement de l'entité sélectionnée).

\subsubsection{Variables indépendantes de la requête et de l'entité sélectionnée}

\begin{itemize}
\item (LEN) La longueur de la phrase.
\item (SSS) Le nombre de parties de la phrase séparées par des signes de ponctuation et d'une longueur inférieure à quatre mots.
\item (SS,SE) La nature du premier caractère (alphabétique, numérique, autre) et du dernier caractère (point, points de suspension, etc.) de la phrase.
\item (HL) La présence d'hyperliens dans la phrase.
\item (D) La présence d'une date dans la phrase.
\item (SL) La position de la phrase dans la page web, normalisée par le nombre total de phrases.
\item (LS) Est-ce la dernière phrase ? (variable binaire)
\item (NCH) Le ratio des caractères non-alphabétiques sur les caractères alphabétiques au sein de la phrase.
\end{itemize}

\subsubsection{Variables dépendantes de la requête et indépendantes de l'entité sélectionnée}

\begin{itemize}
\item (QT) Le nombre de termes partagés par la requête et la phrase.
\item (QR) Le nombre d'entités annotées partagées par la requête et la phrase.
\item (KTKR) Le nombre d'entités classées parmi les top-k par LDRANK et présentes dans la phrase.
\end{itemize}

\subsubsection{Variables dépendantes de l'entité sélectionnée}

\begin{itemize}
\item (SEL) Le nombre de termes partagés par le label de l'entité et la phrase.
\item (SEA) Le nombre de termes partagés par le résumé de l'entité et la phrase.
\item (SEN) Le nombre d'entités voisines de l'entité sélectionnée et présentes dans la phrase.
\item (SENL) Le nombre de termes partagés par les labels des entités voisines de l'entité sélectionnée et la phrase.
\item (SEP) La position relative de l'entité dans la phrase (normalisée par la taille de la phrase).
\item (SES) Le score LDRANK de l'entité sélectionnée (normalisé par le score LDRANK max obtenu par une entité de la même page).
\end{itemize}

\subsubsection{Variables dépendantes des entités annotées dans la phrase}

\begin{itemize}
\item (AE) Le nombre d'entités annotées dans la phrase.
\item (TKE) Le nombre d'entités classées parmi les $k$ premières par LDRANK.
\item (AET) Le nombre de termes partagés par la phrase et les labels des entités annotées dans la phrase.
\item (TEL) Le nombre de termes partagés par la phrase et les labels des $k$ meilleures entités au sens de LDRANK.
\item (AES) Le score LDRANK moyen des entités présentes dans la phrase (normalisé par le score LDRANK max obtenu par une entité de la même page).
\item (AEL) Pour le graphe RDF construit à partir de la page, le nombre d'arêtes reliant des entités annotées dans la phrase.
\end{itemize}

\subsection{Construction du jeu de données}\label{mldataset}

Pour le jeu d'apprentissage, nous avons utilisé celui dont la construction par crowdsourcing a été décrite en section~\ref{dataset}. Il nous suffit d'interpréter ce dernier selon un nouveau point de vue~: ``la pertinence d'une entité dans le contexte d'une phrase et d'un besoin d'information exprimé par une requête'' devient ``la propension pour une phrase de mettre en avant la relation entre une entité et le besoin d'information''. Par ailleurs, nous procédons à une agrégation des jugements pour obtenir un jeu de données à deux classes~: le score 0 (i.e., \textit{irrelevant}) signifie que la phrase n'est pas sélectionnée, les autres scores (1 pour ``marginally relevant'', 2 pour ``fairly relevant'', et 3 pour ``highly relevant'') signifient tous que la phrase est sélectionnée. Remarquons que nous avons expérimenté avec d'autres manières d'agréger les jugements pour obtenir deux classes, mais c'est avec celle décrite ci-dessus que nous avons obtenu les meilleurs résultats expérimentaux.

\subsection{Équilibrage du jeu de données d'apprentissage}\label{mlunbalanced}

Le jeu d'apprentissage comprend 15\,841 instances d'entités détectées dans les pages web. 9\,936 d'entre elles appartiennent à des phrases bien adaptées pour illustrer la relation les liant au besoin d'information, tandis que 5\,905 d'entre elles apparaissent dans des phrases qui ne sont pas pertinentes. Ainsi, la classe positive représente $62,7\,\%$ du nombre totale d'instances du jeu d'apprentissage. Nous remédions à ce léger déséquilibre en appliquant l'algorithme SMOTE (Synthetic Minority Over-Sampling Technique)~\cite{chawla2002smote} qui permet de faire croître la classe minoritaire grâce à des données synthétiques, et également de réduire la taille de la classe majoritaire.

\subsection{Premiers résultats}\label{mlresults1}

Nous avons expérimenté avec cinq algorithmes d'apprentissage~: régression logistique, naive Bayes (avec une distribution Gaussienne comme distribution conditionnelle pour les attributs numériques), l'implémentation J48 de l'algorithme d'arbre de décision \textit{C4.5}, Radial Basis Function Network, et SVM (implémentation libSVM, avec un noyau de type Radial Basis Function pour la version nu-SVC de l'algorithme, et avec $0.5$ pour valeur du paramètre nu qui fixe une borne inférieure sur la fraction des erreurs de classification dues à une marge trop large). Nous avons adopté une approche par validation croisée sur 10 échantillons. Par ailleurs, nous avons utilisé la plateforme Weka~\cite{hall2009weka}. Nous avons utilisé pour métrique d'évaluation le F1-score pour la classe positive (F-True), pour l'ensemble des deux classes (F-All), et l'aire sous la courbe ROC. Le tableau~\ref{tab:mlresults1} présente les résultats de cette expérimentation.

\begin{table}[tab:mlresults1]{Résultats pour l'apprentissage de la sélection de phrases explicatives de la relation entre une entité et le besoin d'information}
$\begin{array}{c|ccc}
& \text{F-True} & \text{F-All} & \text{ROC}\\
\hline
\textbf{Logistic Regression} & \textbf{0,802} & \textbf{0,588} & \textbf{0,658}\\
\text{Naive Bayes} & 0,759 & 0,633 & 0,644\\
\text{J48} & 0,772 & 0,659 & 0,652\\
\text{LIBSVM} & 0,802 & 0,537 & 0,588\\
\text{RBF} & 0,802 & 0,584 & 0,631
\end{array}$
\end{table}

\subsection{Stratégies pour la sélection de variables}\label{mlfeatureselection}

Après l'obtention de ces premiers résultats, nous aimerions réduire le nombre de variables nécessaires sans détériorer la qualité de la prédiction. Nous sommes intéressés par cette réduction du nombre de variables à plusieurs titres. D'abord, avec moins de variables l'exécution de l'algorithme de prédiction consommera moins de mémoire et pourra être plus rapide. Ensuite, du temps d'exécution au moment de la prédiction pourra également être gagné car le calcul des valeurs des variables peut parfois être coûteux. Aussi, contraindre les algorithmes de prédiction à ne pouvoir utiliser qu'un petit sous-ensemble des variables initiales peut réduire les erreurs d'estimation et éviter le sur-apprentissage. Enfin, nous sommes dans notre cas particulièrement intéressés par vérifier indirectement l'efficacité de l'algorithme LDRANK en déterminant l'importance des variables qui font appel à son ordonnancement des entités.

Nous avons d'abord utilisé une approche de sélection de variables par filtre. Il s'agit de filtrer les variables en amont du processus d'apprentissage supervisé. Ainsi, nous avons employé la métrique du gain d'information pour ordonner a priori les variables. Nous avons essayé deux stratégies~: (i) conserver seulement les 10 meilleures variables, (ii) couper au niveau d'un saut net (\textit{nosedive}) dans l'ordonnancement des variables par leur score de gain d'information (ce qui nous a amené à conserver 16 variables, voir la ligne ``Infogain (nosedive)'' du tableau~\ref{tab:mlresults2}).

Nous avons ensuite utilisé une approche enveloppe pour la sélection de variables. Dans ce cas, la sélection utilise la méthode d'apprentissage comme une boîte noire et cherche à optimiser explicitement le taux d'erreur. Il s'agit donc, \textit{a priori}, d'explorer tout l'espace des sous-ensembles de variables de l'ensemble initial, et pour chaque point de cet espace d'exécuter l'apprentissage de l'algorithme de prédiction considéré. L'avantage de ce type d'approche est d'être adapté à l'algorithme d'apprentissage et de prendre en compte des dépendances entre variables. Mais, le coût d'une telle approche appliquée naïvement serait prohibitif. Ainsi, cet espace de recherche sera souvent exploré grâce à une heuristique gloutonne. Deux stratégies classiques sont à considérer~: (i) le mode \textit{avant} (\textit{forward selection}) pour lequel on part d'un petit ensemble de variables que l'on fait croître progressivement, et (ii) le mode \textit{arrière} (\textit{backward elimination}) pour lequel, partant avec toutes les variables, on élimine une à une les plus faibles. Nous avons testé une stratégie gloutonne naïve (descente de gradient sans retour arrière, voir la ligne ``Wrapper back. greedy'' du tableau~\ref{tab:mlresults2}), et une stratégie gloutonne avec possibilité de retour arrière en autorisant le choix d'au plus deux nœuds consécutifs de l'espace de recherche n'améliorant pas le taux d'erreur (voir la ligne ``Wrapper back. best-first'' du tableau~\ref{tab:mlresults2}).

\begin{table}[tab:mlresults2]{Résultats de la sélection de variables pour l'approche par filtre avec métrique InfoGain, et pour l'approche par enveloppe avec deux stratégies d'exploration gloutonne de l'espace des sous-ensembles des variables (les valeurs représentent le score F1 pour la classe positive (F1-True))}
$\begin{array}{c|ccccc}
& \text{J48} & \text{N.Bayes} & \text{Logistic} & \text{SVM} & \text{RBF}\\
\hline
\text{Toutes les var.} & 0,772 & 0,759 & \textbf{0,802} & 0,802 & 0,802\\
\text{Infogain (top 10)} & 0,799 & 0,785 & \textbf{0,803} & 0,799 & 0,801\\
\text{Infogain (nosedive)} & 0,792 & 0,771 & \textbf{0,803} & 0,801 & 0,801\\
\text{Wrapper back. best-first} & 0,772 & 0,802 & \textbf{0,804} & 0,802 & 0,801\\
\text{Wrapper back. greedy} & 0,787 & 0,802 & \textbf{0,802} & 0,802 & 0,802
\end{array}$
\end{table}

Nous avons remarqué que, dans notre cas, l'algorithme de régression logistique était celui qui se comportait le mieux pour la sélection de variables. Ainsi, nous avons essayé d'appliquer pour cet algorithme une approche enveloppe en mode avant en faisant varier le nombre de variables initiales de 0 à 10 dans l'ordre de leur ordonnancement par la métrique de gain d'information (voir le tableau~\ref{tab:mlresults3}).

\begin{table}[tab:mlresults3]{Résultats de la sélection de variables pour l'approche enveloppe en mode avant avec l'algorithme de régression logistique et en faisant varier le nombre de variables initiales de 0 à 10 selon leur score de gain d'information}
$\begin{array}{c|cccc}
\text{Nb var init.} & 10 & \textbf{6} & 4\\
\hline
\text{Nb var finales} & 16 & \textbf{13} & 19\\
\text{F-True} & 0,804 & \textbf{0,804} & 0,803\\
\text{F-All} & 0,588 & \textbf{0,587} & 0,589\\
\text{ROC} & 0,658 & \textbf{0,658} & 0,659
\end{array}$
\end{table}

\subsection{Analyse des résultats}\label{mlresults2}

Par cette stratégie de sélection de variables, nous parvenons finalement à réduire le nombre de variables à 13~:
\begin{itemize}
\item (TEL) Le nombre de termes partagés par la phrase et les labels des $k$ meilleures entités au sens de LDRANK.
\item (SES) Le score LDRANK de l'entité sélectionnée (normalisé par le score LDRANK max obtenu par une entité de la même page).
\item (SEL) Le nombre de termes partagés par le label de l'entité et la phrase.
\item (SEA) Le nombre de termes partagés par le résumé de l'entité et la phrase.
\item (SEN) Le nombre d'entités voisines de l'entité sélectionnée et présentes dans la phrase.
\item (SENL) Le nombre de termes partagés par les labels des entités voisines de l'entité sélectionnée et la phrase.
\item (SEP) La position relative de l'entité dans la phrase (normalisée par la taille de la phrase).
\item (SE) La nature du dernier caractère (point, points de suspension, etc.) de la phrase.
\item (QT) Le nombre de termes partagés par la requête et la phrase.
\item (AE) Le nombre d'entités annotées dans la phrase.
\item (AET) Le nombre de termes partagés par la phrase et les labels des entités annotées dans la phrase.
\item (SL) La position de la phrase dans la page web, normalisée par le nombre total de phrases.
\item (LS) Est-ce la dernière phrase ? (variable binaire)
\end{itemize}

Nous remarquons que les variables construites sur LDRANK, ainsi que celles qui dépendent d'une connaissance extérieure apportée par les entités sont toutes présentes.

\subsection{Synthèse}\label{mlsynthese}

En conclusion, nous avons choisi des variables basées sur la requête, le texte de la page web, et l'ordonnancement des entités produit par LDRANK. Nous avons mis en place un processus de sélection de variables~: nous avons utilisé une métrique de gain d'information pour sélectionner un petit ensemble de variables que nous avons ensuite utilisé comme ensemble de départ pour une approche \textit{enveloppe} (en mode \textit{avant}). Nous avons observé que les variables dérivées de l'ordonnancement produit par le LDRANK appartiennent systématiquement aux variables conservées. Ceci nous semble fournir une preuve supplémentaire, même si indirecte, de l'utilité du LDRANK.

Dans la section suivante, nous décrivons les résultats d'une évaluation par crowdsourcing du système ENsEN.

\section{Évaluation par crowdsourcing de ENsEN}\label{enseneval}

\subsection{Méthodologie}

Nous cherchons à évaluer l'efficacité et l'utilité de notre système de génération de snippets sémantiques, ENsEN, en le comparant à une interface de recherche traditionnelle, celle du moteur de recherche Google. Nous adoptons une approche par crowdsourcing grâce à la plateforme CrowdFlower (voir la section~\ref{dataset} pour une description de cette plateforme).

La tâche centrale d'un participant à notre évaluation est de trouver les réponses à des questions de recherche d'information sur le web (par exemple, ``What division (weight) did the boxer Floyd Patterson win?''). Pour un participant, une première moitié de ses tâches doit être réalisée en utilisant Google, l'autre moitié en utilisant ENsEN. Ensuite, nous demandons à chaque participant d'évaluer et de comparer les deux interfaces.

Chaque tâche est extraite du jeu de données TREC 2004 QA. Nous avons choisi aléatoirement 8 thèmes (\textit{topics}). Chacun est accompagné de trois questions pour lesquelles nous connaissons les réponses. Deux des questions sont factuelles, la troisième attend une réponse sous la forme d'une liste. Par ailleurs, une tâche comprend également un court questionnaire servant à déterminer l'expertise du participant pour le domaine de la recherche sur le web (avec des questions du type~: ``Pour trouver les pages qui comprennent une phrase exacte, faut-il~: (i) entourer la phrase de guillemets, (ii) simplement saisir la phrase telle quelle, (iii) placer un astérisque à une extrémité de la phrase ?~; voir figure~\ref{fig:ensengoogleexpertise} pour les résultats de cette enquête). Enfin, nous collectons quelques données sur l'âge des participants et leurs habitudes de recherche~: font-ils le plus souvent des requêtes exploratoires, informationnelles, navigationnelles ou transactionnelles ? Voir les figures~\ref{fig:ensenevalages} et~\ref{fig:ensentypequeries} pour les résultats de cette enquête. 

\begin{figure}[fig:ensengoogleexpertise]{Expertise des participants pour le domaine de la recherche sur le web en fonction de la précision de leurs réponses à un questionnaire}
\includegraphics[width=4.9in, keepaspectratio]{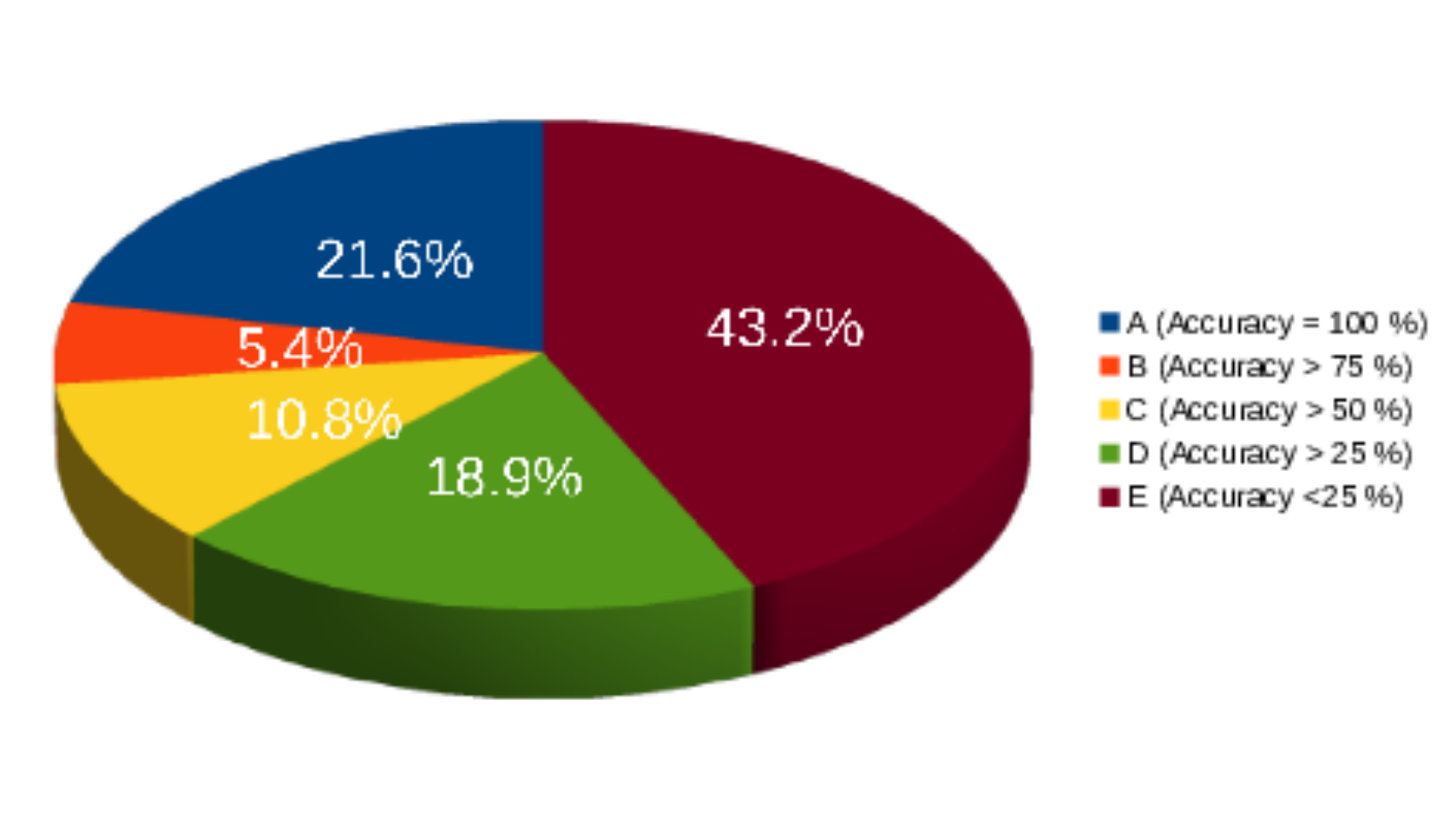}
\end{figure}

\begin{figure}[fig:ensenevalages]{Ages des participants à l'évaluation de ENsEN}
\includegraphics[width=4.9in, keepaspectratio]{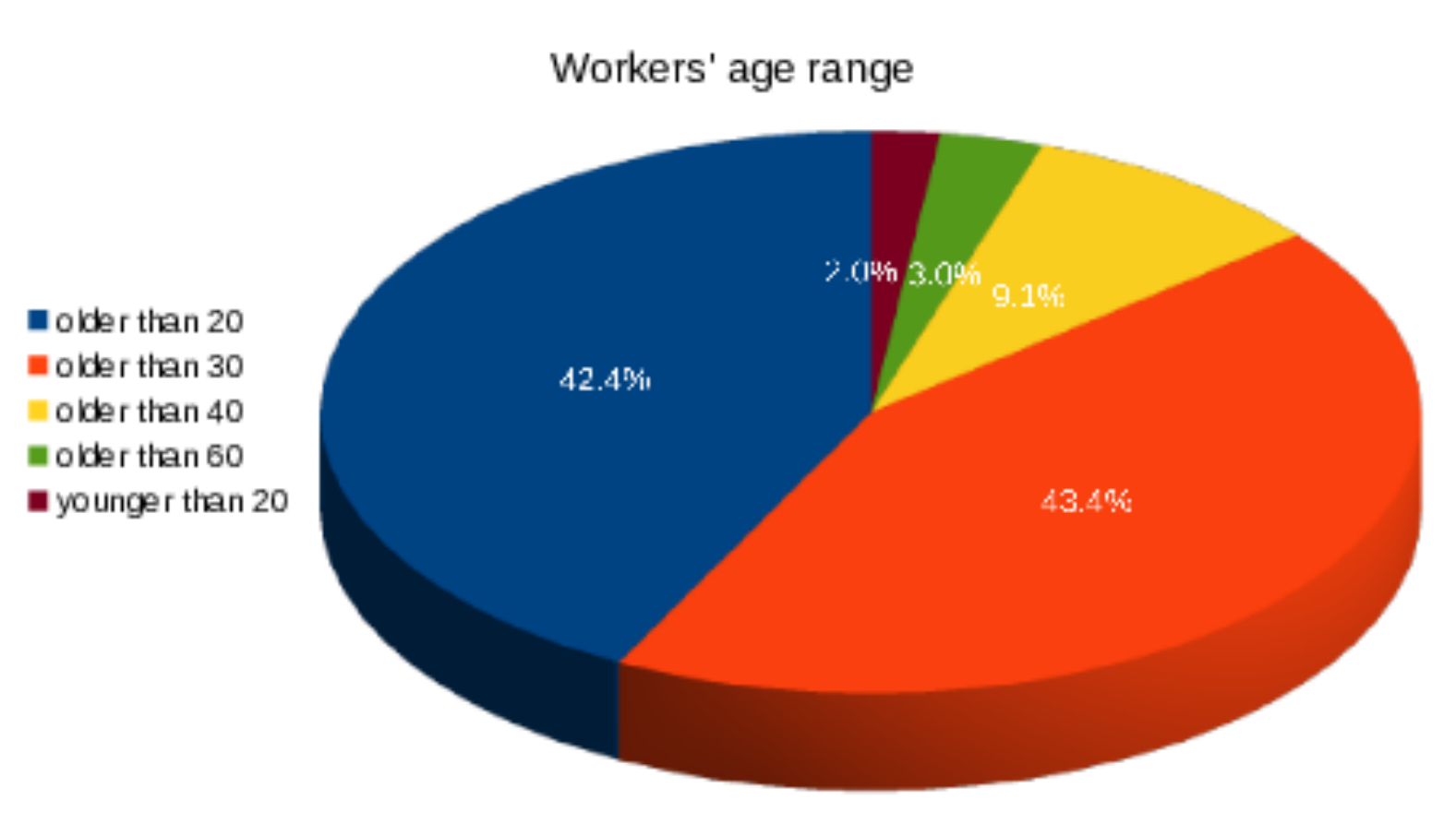}
\end{figure}

\begin{figure}[fig:ensentypequeries]{Habitudes de recherche des participants en termes des types de requêtes qu'ils font le plus souvent}
\includegraphics[width=4.9in, keepaspectratio]{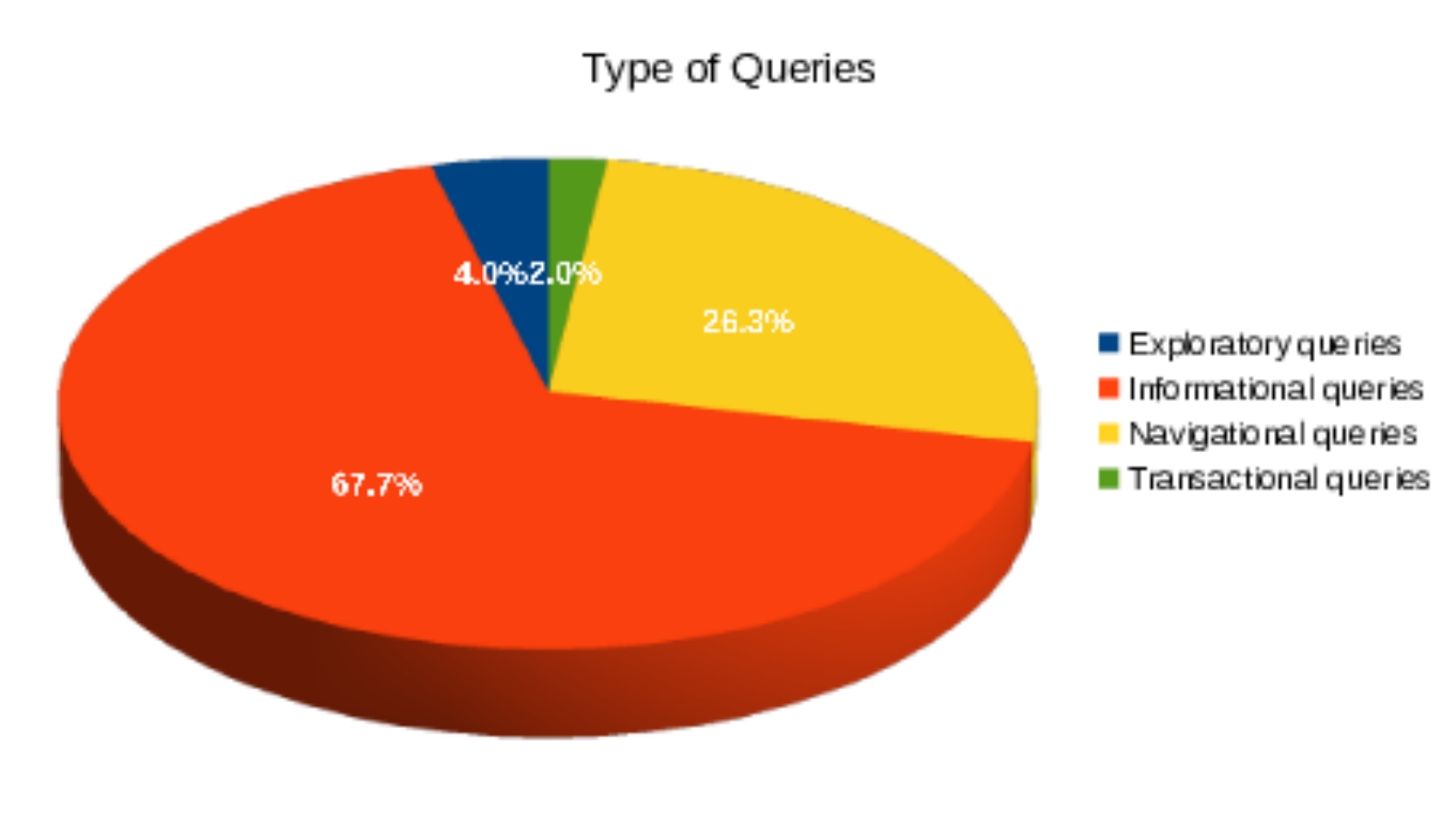}
\end{figure}

Nous avons obtenu 11 jugements pour chacun des 8 thèmes. Nous n'acceptons que les jugements dont la réponse a été fournie en moins de 30~min. Nous n'avons gardé que les meilleurs participants du point de vue d'une métrique propre à CrowdFlower. Nous décrivions brièvement cette dernière métrique en section~\ref{dataset}, elle se fonde sur l'exploitation de tâches tests dont les réponses ont été fournies par le concepteur de la tâche au système CrowdFlower. Ainsi, pour chacun des 8 thèmes, nous générons une tâche test pour laquelle le participant doit répondre en utilisant le moteur de recherche Google.

Avant de pouvoir être rémunéré en répondant à des questions, un participant doit passer par le mode dit ``quiz''. A cette occasion, une tâche test lui est présentée. S'il fait plus de 50\,\% d'erreurs à ce test, il se verra interdire l'accès au mode normal rémunéré. En mode normal, un participant se voit présenter une page avec deux tâches~: l'une demande d'utiliser Google, l'autre ENsEN. Le travail nécessaire pour compléter une telle page faite de deux tâches est rémunéré \$0.25.

\subsection{Analyse des résultats}

Tout d'abord, nous remarquons que la précision des réponse ne dépend pas du niveau d'expertise des participants pour le domaine de la recherche sur le web (voir figure~\ref{fig:ensenaccuracyperclass}).

\begin{figure}[fig:ensenaccuracyperclass]{Précision des réponses en fonction du niveau d'expertise des participants pour le domaine de la recherche sur le web}
\includegraphics[width=4.9in, keepaspectratio]{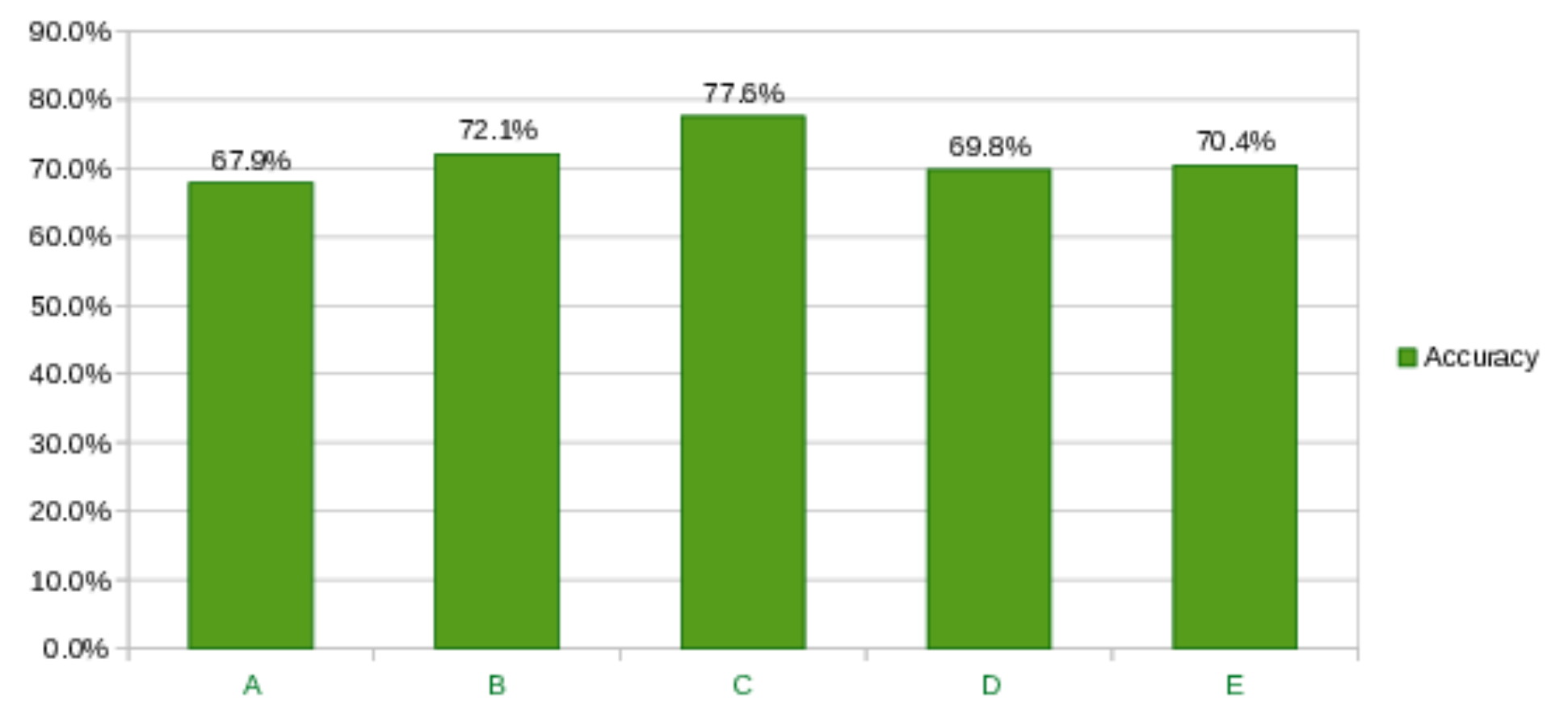}
\end{figure}

Par ailleurs, nous observons que la précision des réponses est le plus souvent comparable pour les deux système Google et ENsEN, sauf dans le cas des thèmes pour lesquels la précision des réponses est la moins bonne (c'est-à-dire les thèmes 1 et 7 qui sont sans doute plus difficiles) où ENsEN est significativement meilleur que Google (voir figure~\ref{fig:ensenaccuracypertopicandsystem}).

\begin{figure}[fig:ensenaccuracypertopicandsystem]{Précision des réponses par thème et pour chacun des deux systèmes}
\includegraphics[width=4.9in, keepaspectratio]{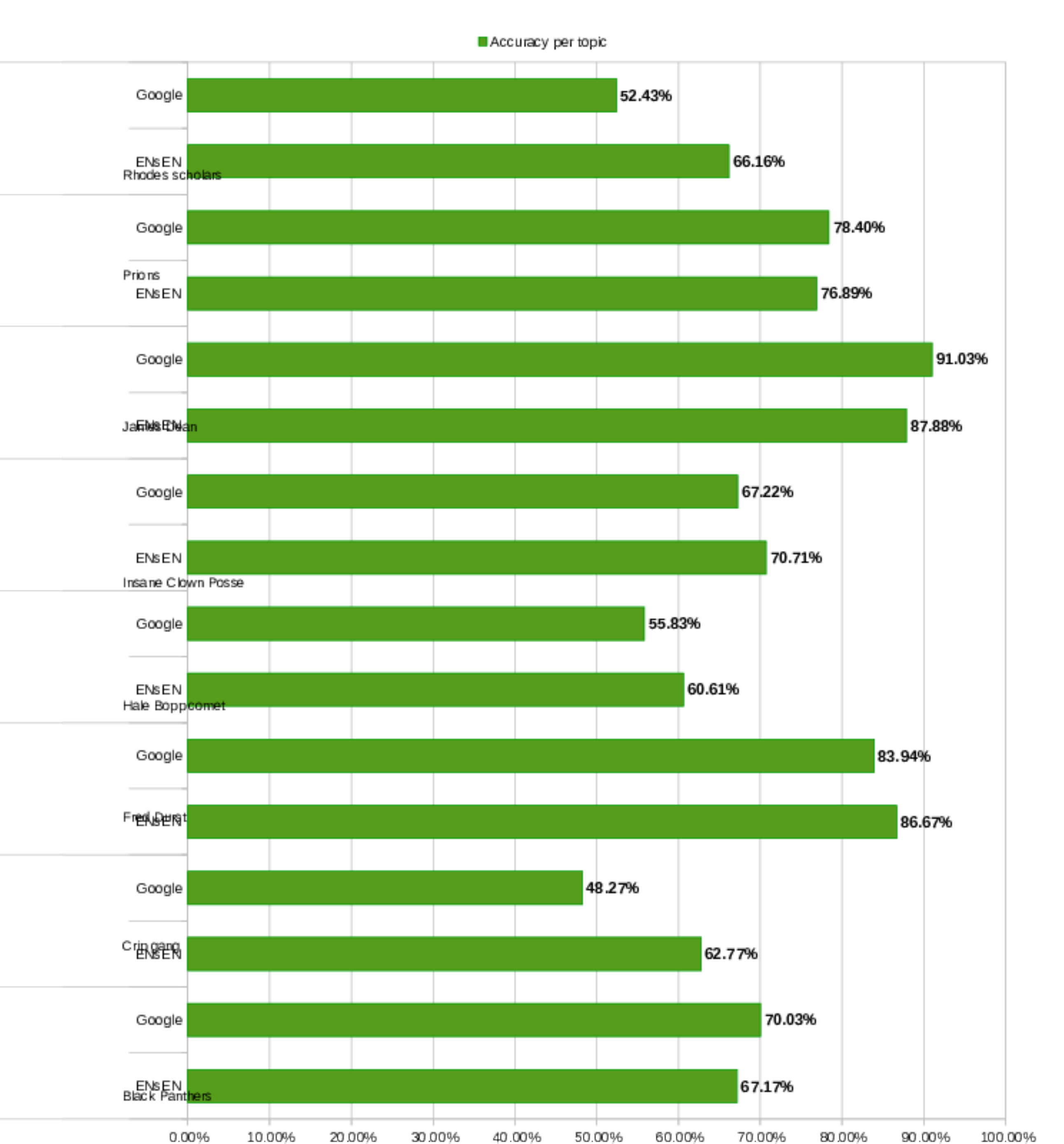}
\end{figure}

Aussi, pour les deux types de questions (c'est-à-dire, (i) donner la bonne réponse, (ii) donner une liste de bonnes réponses), nous remarquons que si les utilisateurs préfèrent l'interface homme-machine proposée par Google, ils ont cependant l'impression de trouver plus facilement la réponse correcte en utilisant ENsEN (voir figures~\ref{fig:ensenoverallq1} et~\ref{fig:ensenoverallq2}).

\begin{figure}[fig:ensenoverallq1]{Évaluation globale de l'efficacité et de la facilité d'utilisation pour les questions factuelles}
\includegraphics[width=4.9in, keepaspectratio]{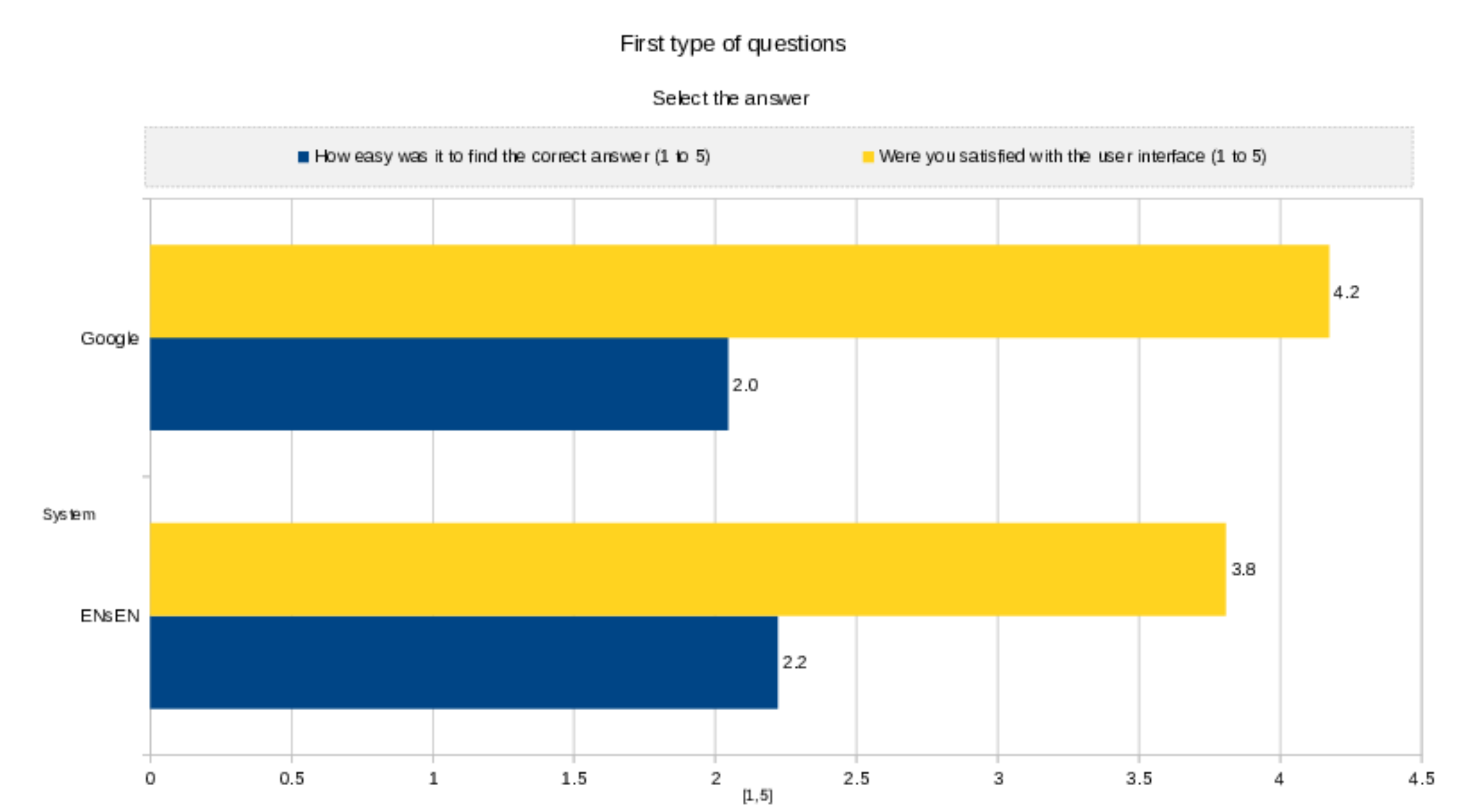}
\end{figure}

\begin{figure}[fig:ensenoverallq2]{Évaluation globale de l'efficacité et de la facilité d'utilisation pour les questions dont la réponse est une liste}
\includegraphics[width=4.9in, keepaspectratio]{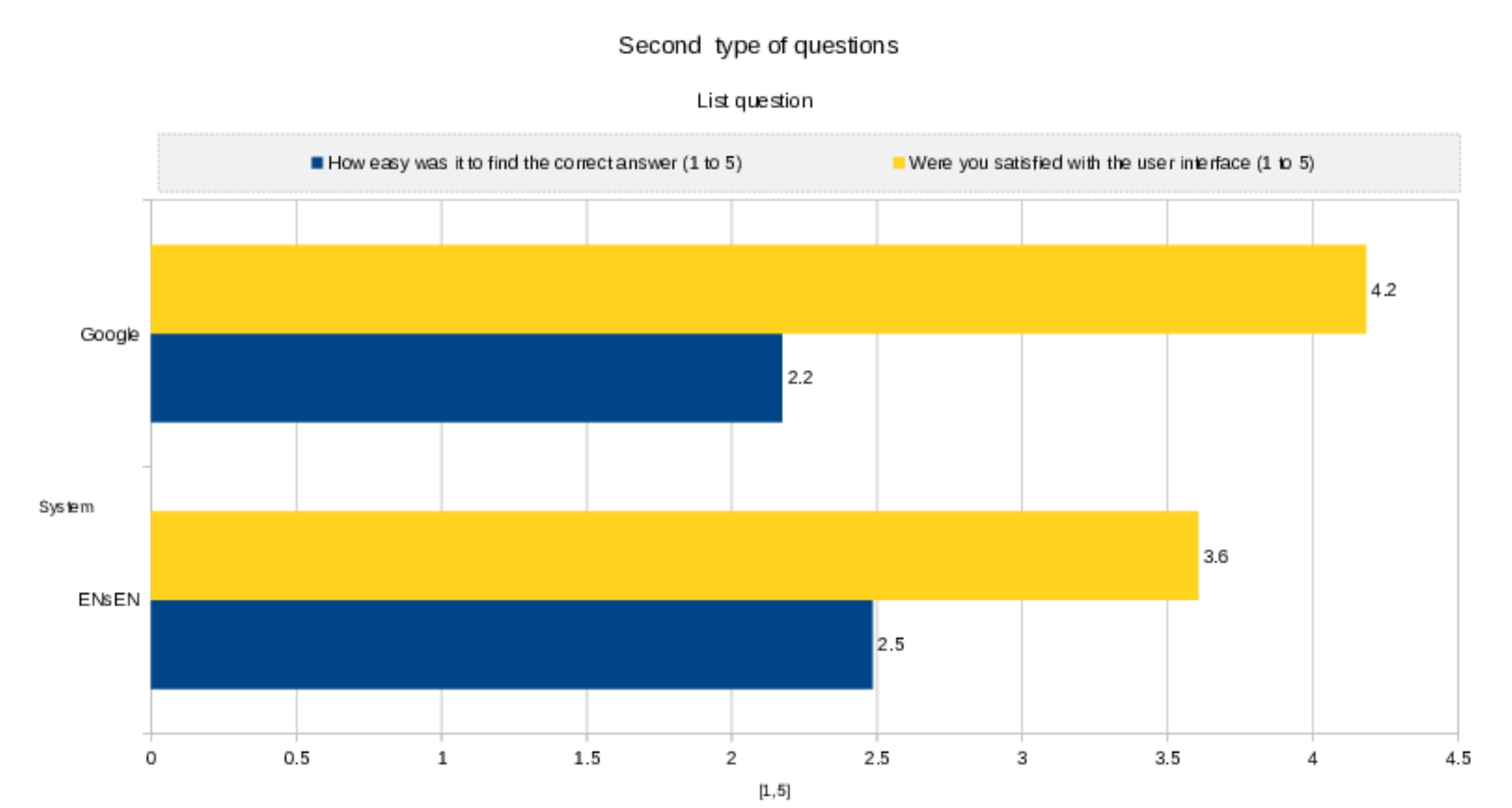}
\end{figure}

Enfin, lorsque nous demandons aux participants quels sont les éléments de l'interface d'ENsEN qui leur semblent les plus utiles, nous observons que les parties de l'interface qui mettent en avant les meilleures entités découvertes par LDRANK sont très appréciées (voir figure~\ref{fig:ensenimportantparts}).

\begin{figure}[fig:ensenimportantparts]{Éléments de l'interface préférés par les participants}
\includegraphics[width=4.9in, keepaspectratio]{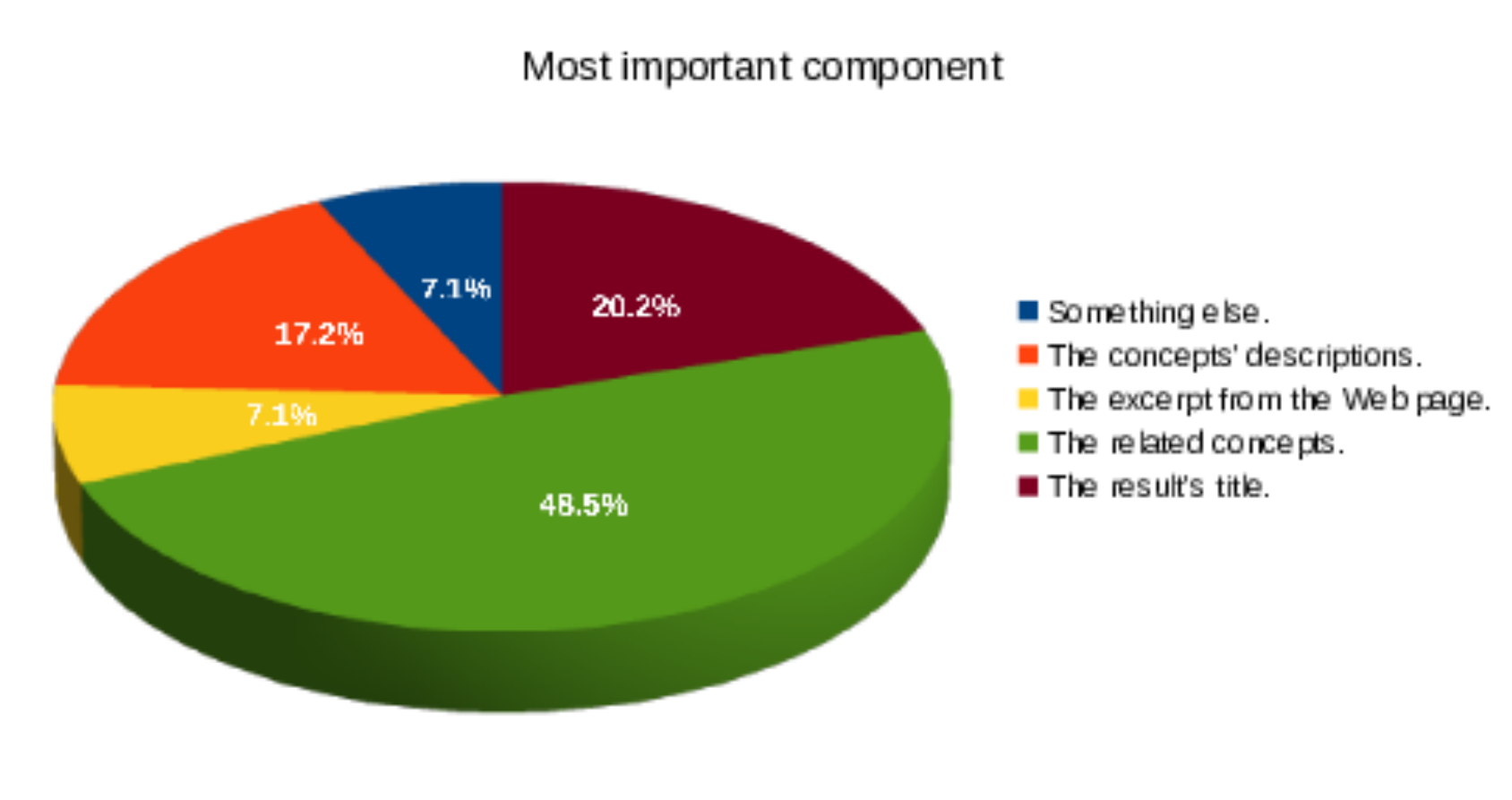}
\end{figure}

\section{Conclusion}\label{conclusion}

Nous avons proposé un nouvel algorithme, LDRANK, pour l'ordonnancement des entités d'un graphe du web des données qui peut être creux et bruité, mais pour lequel des données textuelles descriptives sont associées aux n{\oe}uds, et avec connaissance d'un besoin d'information exprimé sous la forme d'un ensemble de mots clés. De tels graphes apparaissent en particulier suite à un processus de détection automatique d'entités du web des données au sein d'une page web (par exemple, à travers l'utilisation de DBpedia Spotlight). Notre approche se caractérise par la prise en compte à la fois de la structure explicite offerte par le web des données, et des relations implicites qui peuvent être trouvées par analyse du texte de la page web.

LDRANK construit des ordonnancements d'une qualité significativement meilleure à celle des ordonnancements produits par les approches de l'état de l'art. De plus, nous avons appliqué cet algorithme au cadre applicatif de la construction de snippets sémantiques. Dans ce contexte, la bonne précision de l'algorithme LDRANK a permis d'obtenir des snippets sémantiques utiles et utilisables, ouvrant la voie à de nouvelles applications qui pourront bénéficier des apports mutuels du web des données et du web des documents. Par exemple, des travaux futurs évalueront le potentiel de cette approche pour la recherche d'information exploratoire.

\bibliography{dn2015}

\end{document}